\documentclass[apj,twocolappendix]{emulateapj}

\usepackage{amsmath}
\usepackage{verbatim} 
\usepackage{enumerate}
\usepackage{appendix}
\usepackage{listings}
\usepackage{mathrsfs}
\usepackage{standalone}
\usepackage[flushleft]{threeparttable}
\usepackage{adjustbox}

\newcommand{\bjdtdb}{\ensuremath{\rm {BJD_{TDB}}}}

\newcommand{\jdutc}{\ensuremath{\rm {JD_{UTC}}}}

\newcommand{\ms}{\ensuremath{M_*}}
\newcommand{\rs}{\ensuremath{R_*}}

\newcommand{\fave}{\langle F \rangle}
\newcommand{\fluxcgs}{10$^9$ erg s$^{-1}$ cm$^{-2}$}

\newcommand{\ecosw}{\ensuremath{e\cos{\omega_*}}}
\newcommand{\esinw}{\ensuremath{e\sin{\omega_*}}}

\newcommand{\vsini}{\ensuremath{v\sin{I_*}}}

\newcommand{\logg}{\ensuremath{\log g}}

\newcommand{\feh}{\ensuremath{\left[{\rm Fe}/{\rm H}\right]}}
\newcommand{\teff}{\ensuremath{T_{\rm eff}}}

\newcommand{\msun}{\ensuremath{\,{\rm M}_\Sun}}
\newcommand{\rsun}{\ensuremath{\,{\rm R}_\Sun}}
\newcommand{\lsun}{\ensuremath{\,{\rm L}_\Sun}}
\newcommand{\mj}{\ensuremath{\,{\rm M}_{\rm J}}}
\newcommand{\rj}{\ensuremath{\,{\rm R}_{\rm J}}}

\newcommand{\rchisq}{\ensuremath{\chi_\nu^{\,2}}}
\newcommand{\chisq}{\ensuremath{\chi^{\,2}}}

\newcommand{\starname}{KELT-4A}
\newcommand{\pname}{\starname \lowercase{b}}

\newcommand{\mstarvalone}{\ensuremath{1.204_{-0.063}^{+0.072}}}
\newcommand{\mstarvaltwo}{\ensuremath{1.201_{-0.061}^{+0.067}}}
\newcommand{\rstarvalone}{\ensuremath{1.610_{-0.068}^{+0.078}}}
\newcommand{\rstarvaltwo}{\ensuremath{1.603_{-0.038}^{+0.039}}}
\newcommand{\lstarvalone}{\ensuremath{3.46_{-0.38}^{+0.43}}}
\newcommand{\lstarvaltwo}{\ensuremath{3.43_{-0.27}^{+0.28}}}
\newcommand{\rhovalone}{\ensuremath{0.407\pm0.044}}
\newcommand{\rhovaltwo}{\ensuremath{0.411_{-0.017}^{+0.018}}}
\newcommand{\agevalone}{\ensuremath{4.38_{-0.88}^{+0.81}}}
\newcommand{\agevaltwo}{\ensuremath{4.44_{-0.89}^{+0.78}}}
\newcommand{\loggstarvalone}{\ensuremath{4.105_{-0.032}^{+0.029}}}
\newcommand{\loggstarvaltwo}{\ensuremath{4.108\pm0.014}}
\newcommand{\teffvalone}{\ensuremath{6207_{-76}^{+75}}}
\newcommand{\teffvaltwo}{\ensuremath{6206\pm75}}
\newcommand{\fehvalone}{\ensuremath{-0.116_{-0.071}^{+0.067}}}
\newcommand{\fehvaltwo}{\ensuremath{-0.116_{-0.069}^{+0.065}}}
\newcommand{\vsinistarvalone}{\ensuremath{6000\pm1200}}
\newcommand{\vsinistarvaltwo}{\ensuremath{6000\pm1200}}
\newcommand{\lambdavalone}{\ensuremath{30_{-74}^{+85}}}
\newcommand{\lambdavaltwo}{\ensuremath{14_{-64}^{+100}}}
\newcommand{\dvalone}{\ensuremath{211_{-12}^{+13}}}
\newcommand{\dvaltwo}{\ensuremath{210.7\pm9.0}}
\newcommand{\evalone}{\ensuremath{0.030_{-0.021}^{+0.036}}}
\newcommand{\evaltwo}{\ensuremath{--}}
\newcommand{\omegavalone}{\ensuremath{60_{-120}^{+110}}}
\newcommand{\omegavaltwo}{\ensuremath{--}}
\newcommand{\pvalone}{\ensuremath{2.9895933\pm0.0000049}}
\newcommand{\pvaltwo}{\ensuremath{2.9895932\pm0.0000049}}
\newcommand{\avalone}{\ensuremath{0.04321_{-0.00077}^{+0.00085}}}
\newcommand{\avaltwo}{\ensuremath{0.04317_{-0.00074}^{+0.00079}}}
\newcommand{\mpvalone}{\ensuremath{0.878_{-0.067}^{+0.070}}}
\newcommand{\mpvaltwo}{\ensuremath{0.902_{-0.059}^{+0.060}}}
\newcommand{\rpvalone}{\ensuremath{1.706_{-0.076}^{+0.085}}}
\newcommand{\rpvaltwo}{\ensuremath{1.699_{-0.045}^{+0.046}}}
\newcommand{\rhopvalone}{\ensuremath{0.219_{-0.029}^{+0.031}}}
\newcommand{\rhopvaltwo}{\ensuremath{0.228_{-0.018}^{+0.019}}}
\newcommand{\loggpvalone}{\ensuremath{2.873_{-0.045}^{+0.042}}}
\newcommand{\loggpvaltwo}{\ensuremath{2.889_{-0.030}^{+0.029}}}
\newcommand{\teqvalone}{\ensuremath{1827_{-42}^{+44}}}
\newcommand{\teqvaltwo}{\ensuremath{1823\pm27}}
\newcommand{\thetavalone}{\ensuremath{0.0368_{-0.0029}^{+0.0030}}}
\newcommand{\thetavaltwo}{\ensuremath{0.0381\pm0.0024}}
\newcommand{\favevalone}{\ensuremath{2.53_{-0.23}^{+0.25}}}
\newcommand{\favevaltwo}{\ensuremath{2.51\pm0.15}}
\newcommand{\tcvalone}{\ensuremath{2456190.30201\pm0.00022}}
\newcommand{\tcvaltwo}{\ensuremath{2456190.30201\pm0.00022}}
\newcommand{\tpvalone}{\ensuremath{2456190.09_{-0.97}^{+0.85}}}
\newcommand{\tpvaltwo}{\ensuremath{--}}
\newcommand{\kvalone}{\ensuremath{108.6\pm7.4}}
\newcommand{\kvaltwo}{\ensuremath{111.8_{-6.4}^{+6.3}}}
\newcommand{\krvalone}{\ensuremath{72\pm14}}
\newcommand{\krvaltwo}{\ensuremath{71\pm14}}
\newcommand{\mpsinivalone}{\ensuremath{0.871_{-0.066}^{+0.069}}}
\newcommand{\mpsinivaltwo}{\ensuremath{0.896_{-0.058}^{+0.060}}}
\newcommand{\mpmstarvalone}{\ensuremath{0.000695\pm0.000049}}
\newcommand{\mpmstarvaltwo}{\ensuremath{0.000717\pm0.000042}}
\newcommand{\uvalone}{\ensuremath{0.6018_{-0.0068}^{+0.0078}}}
\newcommand{\uvaltwo}{\ensuremath{0.6018_{-0.0067}^{+0.0077}}}
\newcommand{\gammaexpertvalone}{\ensuremath{317\pm23}}
\newcommand{\gammaexpertvaltwo}{\ensuremath{317\pm16}}
\newcommand{\gammafiesvalone}{\ensuremath{-98\pm13}}
\newcommand{\gammafiesvaltwo}{\ensuremath{-99\pm12}}
\newcommand{\gammahiresvalone}{\ensuremath{15.7\pm7.4}}
\newcommand{\gammahiresvaltwo}{\ensuremath{15.9\pm6.7}}
\newcommand{\gammatresvalone}{\ensuremath{-11\pm15}}
\newcommand{\gammatresvaltwo}{\ensuremath{-10\pm13}}
\newcommand{\dotgammavalone}{\ensuremath{-0.014\pm0.044}}
\newcommand{\dotgammavaltwo}{\ensuremath{-0.013_{-0.041}^{+0.040}}}
\newcommand{\ecoswvalone}{\ensuremath{0.004_{-0.017}^{+0.025}}}
\newcommand{\ecoswvaltwo}{\ensuremath{--}}
\newcommand{\esinwvalone}{\ensuremath{0.002_{-0.029}^{+0.039}}}
\newcommand{\esinwvaltwo}{\ensuremath{--}}
\newcommand{\fmonemtwovalone}{\ensuremath{0.000000414_{-0.000000079}^{+0.000000091}}}
\newcommand{\fmonemtwovaltwo}{\ensuremath{0.000000454_{-0.000000073}^{+0.000000081}}}
\newcommand{\rprstarvalone}{\ensuremath{0.10892_{-0.00055}^{+0.00054}}}
\newcommand{\rprstarvaltwo}{\ensuremath{0.10893\pm0.00054}}
\newcommand{\arvalone}{\ensuremath{5.77_{-0.22}^{+0.20}}}
\newcommand{\arvaltwo}{\ensuremath{5.792_{-0.082}^{+0.086}}}
\newcommand{\ivalone}{\ensuremath{83.11_{-0.57}^{+0.48}}}
\newcommand{\ivaltwo}{\ensuremath{83.16_{-0.21}^{+0.22}}}
\newcommand{\bvalone}{\ensuremath{0.689_{-0.012}^{+0.011}}}
\newcommand{\bvaltwo}{\ensuremath{0.689_{-0.012}^{+0.011}}}
\newcommand{\deltavalone}{\ensuremath{0.01186\pm0.00012}}
\newcommand{\deltavaltwo}{\ensuremath{0.01187\pm0.00012}}
\newcommand{\tfwhmvalone}{\ensuremath{0.11893\pm0.00045}}
\newcommand{\tfwhmvaltwo}{\ensuremath{0.11892\pm0.00044}}
\newcommand{\tauvalone}{\ensuremath{0.02535_{-0.00089}^{+0.00090}}}
\newcommand{\tauvaltwo}{\ensuremath{0.02536_{-0.00088}^{+0.00089}}}
\newcommand{\tonefourvalone}{\ensuremath{0.14428_{-0.00083}^{+0.00084}}}
\newcommand{\tonefourvaltwo}{\ensuremath{0.14428_{-0.00083}^{+0.00084}}}
\newcommand{\ptvalone}{\ensuremath{0.1548_{-0.0092}^{+0.012}}}
\newcommand{\ptvaltwo}{\ensuremath{0.1539\pm0.0022}}
\newcommand{\ptgvalone}{\ensuremath{0.193_{-0.011}^{+0.015}}}
\newcommand{\ptgvaltwo}{\ensuremath{0.1915_{-0.0029}^{+0.0028}}}

\newcommand{\tczerovaltwo}{\ensuremath{2456016.9045\pm0.0011}}

\newcommand{\tconevaltwo}{\ensuremath{2456025.8724\pm0.0015}}

\newcommand{\tctwovaltwo}{\ensuremath{2456037.8371\pm0.0013}}

\newcommand{\tcthreevaltwo}{\ensuremath{2456046.80185\pm0.00092}}

\newcommand{\tcfourvaltwo}{\ensuremath{2456055.7703_{-0.0013}^{+0.0014}}}

\newcommand{\tcfivevaltwo}{\ensuremath{2456061.7511\pm0.0012}}

\newcommand{\tcsixvaltwo}{\ensuremath{2456070.7183\pm0.0013}}

\newcommand{\tcsevenvaltwo}{\ensuremath{2456088.6563\pm0.0012}}

\newcommand{\tceightvaltwo}{\ensuremath{2456279.98810\pm0.00038}}

\newcommand{\tcninevaltwo}{\ensuremath{2456288.95246_{-0.00089}^{+0.00087}}}

\newcommand{\tconezerovaltwo}{\ensuremath{2456300.91808_{-0.00063}^{+0.00064}}}

\newcommand{\tconeonevaltwo}{\ensuremath{2456312.87644\pm0.00036}}

\newcommand{\tconetwovaltwo}{\ensuremath{2456324.8362_{-0.0013}^{+0.0012}}}

\newcommand{\tconethreevaltwo}{\ensuremath{2456327.82297\pm0.00037}}

\newcommand{\tconefourvaltwo}{\ensuremath{2456333.80597_{-0.00098}^{+0.00096}}}

\newcommand{\tconefivevaltwo}{\ensuremath{2456336.79097_{-0.00099}^{+0.0010}}}

\newcommand{\tconesixvaltwo}{\ensuremath{2456336.79221_{-0.00058}^{+0.00057}}}

\newcommand{\tconesevenvaltwo}{\ensuremath{2456363.69882\pm0.00041}}

\newcommand{\tconeeightvaltwo}{\ensuremath{2456399.5754_{-0.0024}^{+0.0025}}}
\newcommand{\uonecorotvalone}{\ensuremath{0.3216}}
\newcommand{\uonecorotvaltwo}{\ensuremath{0.3217}}
\newcommand{\utwocorotvalone}{\ensuremath{0.2970}}
\newcommand{\utwocorotvaltwo}{\ensuremath{0.2969}}
\newcommand{\uoneivalone}{\ensuremath{0.2216}}
\newcommand{\uoneivaltwo}{\ensuremath{0.2217}}
\newcommand{\utwoivalone}{\ensuremath{0.3022}}
\newcommand{\utwoivaltwo}{\ensuremath{0.3022}}
\newcommand{\uonesloangvalone}{\ensuremath{0.472}}
\newcommand{\uonesloangvaltwo}{\ensuremath{0.472}}
\newcommand{\utwosloangvalone}{\ensuremath{0.2697}}
\newcommand{\utwosloangvaltwo}{\ensuremath{0.2697}}
\newcommand{\uonesloanivalone}{\ensuremath{0.2395}}
\newcommand{\uonesloanivaltwo}{\ensuremath{0.2396}}
\newcommand{\utwosloanivalone}{\ensuremath{0.3045}}
\newcommand{\utwosloanivaltwo}{\ensuremath{0.3044}}
\newcommand{\uonesloanrvalone}{\ensuremath{0.3126}}
\newcommand{\uonesloanrvaltwo}{\ensuremath{0.3127}}
\newcommand{\utwosloanrvalone}{\ensuremath{0.3144}}
\newcommand{\utwosloanrvaltwo}{\ensuremath{0.3143}}
\newcommand{\uonesloanzvalone}{\ensuremath{0.1892}}
\newcommand{\uonesloanzvaltwo}{\ensuremath{0.1893}}
\newcommand{\utwosloanzvalone}{\ensuremath{0.2949}}
\newcommand{\utwosloanzvaltwo}{\ensuremath{0.2948}}
\newcommand{\tsvalone}{\ensuremath{2456188.815_{-0.032}^{+0.047}}}
\newcommand{\tsvaltwo}{\ensuremath{2456188.80721\pm0.00022}}
\newcommand{\bsvalone}{\ensuremath{0.693_{-0.043}^{+0.055}}}
\newcommand{\bsvaltwo}{\ensuremath{--}}
\newcommand{\tsfwhmvalone}{\ensuremath{0.11840_{-0.0027}^{+0.00073}}}
\newcommand{\tsfwhmvaltwo}{\ensuremath{--}}
\newcommand{\tausvalone}{\ensuremath{0.0256_{-0.0029}^{+0.0046}}}
\newcommand{\tausvaltwo}{\ensuremath{--}}
\newcommand{\tsonefourvalone}{\ensuremath{0.1445_{-0.0038}^{+0.0027}}}
\newcommand{\tsonefourvaltwo}{\ensuremath{--}}
\newcommand{\psvalone}{\ensuremath{0.1540\pm0.0022}}
\newcommand{\psvaltwo}{\ensuremath{--}}
\newcommand{\psgvalone}{\ensuremath{0.1917\pm0.0029}}
\newcommand{\psgvaltwo}{\ensuremath{--}}

\newcommand{\ttvaltwo}{\ensuremath{2456193.29157 \pm 0.00021}}
\newcommand{\pervaltwo}{\ensuremath{2.9895936 \pm 0.0000048}}

\shorttitle{\pname} 
\shortauthors{EASTMAN ET AL\@.}

\begin{document}
\title{\pname: An inflated Hot Jupiter transiting the bright ($V\sim10$) component of a hierarchical triple}

\author{
  Jason D. Eastman\altaffilmark{1},
  Thomas G.\ Beatty\altaffilmark{2,3},
  Robert J.\ Siverd\altaffilmark{4},
  Joseph M. O. Antognini\altaffilmark{5},
  Matthew T. Penny\altaffilmark{5,6},
  Erica J. Gonzales\altaffilmark{7},
  Justin R.\ Crepp\altaffilmark{7},
  Andrew W. Howard\altaffilmark{8},
  Ryan L.\ Avril\altaffilmark{9},
  Allyson Bieryla\altaffilmark{1},
  Karen Collins\altaffilmark{10},
  Benjamin J.\ Fulton\altaffilmark{8,11},
  Jian Ge\altaffilmark{12},
  Joao Gregorio\altaffilmark{13},
  Bo Ma\altaffilmark{12},
  Samuel N.\ Mellon\altaffilmark{9,14},
  Thomas E.\ Oberst\altaffilmark{9},
  Ji Wang\altaffilmark{15},
  B.\ Scott Gaudi\altaffilmark{5},
  Joshua Pepper\altaffilmark{16},
  Keivan G.\ Stassun\altaffilmark{10},
  Lars A.\ Buchhave\altaffilmark{1},
  Eric L.\ N.\ Jensen\altaffilmark{17},
  David W.\ Latham\altaffilmark{1},
  Perry Berlind\altaffilmark{1},
  Michael L.\ Calkins\altaffilmark{1},
  Phillip A.\ Cargile\altaffilmark{1},
  Knicole D. Col\'{o}n\altaffilmark{16,18,19},
  Saurav Dhital\altaffilmark{20},
  Gilbert A.\ Esquerdo\altaffilmark{1},
  John Asher Johnson\altaffilmark{1},
  John F.\ Kielkopf\altaffilmark{21},
  Mark Manner\altaffilmark{22},
  Qingqing Mao\altaffilmark{10},
  Kim K. McLeod\altaffilmark{23},
  Kaloyan Penev\altaffilmark{24},
  Robert P.\ Stefanik\altaffilmark{1},
  Rachel Street\altaffilmark{4},
  Roberto Zambelli\altaffilmark{25},
  D. L.\ DePoy\altaffilmark{26},
  Andrew Gould\altaffilmark{5},
  Jennifer L.\ Marshall\altaffilmark{26},
  Richard W.\ Pogge\altaffilmark{5},
  Mark Trueblood\altaffilmark{27},
  Patricia Trueblood\altaffilmark{27}
}
\altaffiltext{1}{Harvard-Smithsonian Center for Astrophysics, Cambridge, MA 02138, USA}
\altaffiltext{2}{Department of Astronomy \& Astrophysics, The Pennsylvania State University, 525 Davey Lab, University Park, PA 16802, USA}
\altaffiltext{3}{Center for Exoplanets and Habitable Worlds, The Pennsylvania State University, 525 Davey Lab, University Park, PA 16802, USA}
\altaffiltext{4}{Las Cumbres Observatory Global Telescope Network, Santa Barbara, CA 93117, USA}
\altaffiltext{5}{Department of Astronomy, The Ohio State University, Columbus, OH 43210, USA}
\altaffiltext{6}{Sagan Fellow}
\altaffiltext{7}{Department of Physics, University of Notre Dame, Notre Dame, IN 46556, USA}
\altaffiltext{8}{Institute for Astronomy, University of Hawaii, Honolulu, HI 96822, USA}
\altaffiltext{9}{Department of Physics, Westminster College, New Wilmington, PA 16172}
\altaffiltext{10}{Department of Physics and Astronomy, Vanderbilt University, Nashville, TN 37235, USA}
\altaffiltext{11}{NSF Graduate Research Fellow}
\altaffiltext{12}{Department of Astronomy, University of Florida, 211 Bryant Space Science Center, Gainesville, FL, 32611}
\altaffiltext{13}{Atalaia Group and Crow-Observatory, Portalegre, Portugal}
\altaffiltext{14}{Department of Physics and Astronomy, University of Rochester, Rochester, NY 14627}
\altaffiltext{15}{Department of Astrophysics, California Institute of Technology, MC 249-17, Pasadena, CA 91125, USA}
\altaffiltext{16}{Department of Physics, Lehigh University, Bethlehem, PA  18015}
\altaffiltext{17}{Department of Physics and Astronomy, Swarthmore College, Swarthmore, PA 19081, USA}
\altaffiltext{18}{NASA Ames Research Center, M/S 244-30, Moffett Field, CA 94035, USA}
\altaffiltext{19}{Bay Area Environmental Research Institute, 625 2nd St. Ste 209 Petaluma, CA 94952, USA}
\altaffiltext{20}{Department of Astronomy, Boston University, 725 Commonwealth Avenue, Boston, MA 02215, USA}
\altaffiltext{21}{Department of Physics \& Astronomy, University of Louisville, Louisville, KY 40292, USA}
\altaffiltext{22}{Spot Observatory, Nunnelly, TN 37137, USA}
\altaffiltext{23}{Wellesley College, Wellesley, MA 02481, USA}
\altaffiltext{24}{Department of Astrophysical Sciences, Princeton University, Peyton Hall, Princeton, NJ 08544, USA}
\altaffiltext{25}{Societ Astronomica Lunae, Castelnuovo Magra 19030, Via Montefrancio, 77, Italy}
\altaffiltext{26}{George P. and Cynthia Woods Mitchell Institute for Fundamental Physics and Astronomy, and Department of Physics and Astronomy, Texas A \& M University, College Station, TX 77843-4242, USA
}
\altaffiltext{27}{Winer Observatory, Sonoita, AZ 85637, USA}

\begin{abstract}

We report the discovery of \pname, an inflated, transiting Hot Jupiter orbiting the brightest component of a hierarchical triple stellar system. The host star is an F star with \teff=\teffvaltwo \ K, \logg=\loggstarvaltwo, \feh=\fehvaltwo, \ms=\mstarvaltwo \ \msun, and \rs=\rstarvaltwo \ \rsun. The best-fit linear ephemeris is $\bjdtdb = \ttvaltwo + E\left(\pervaltwo\right)$. With a magnitude of $V\sim10$, a planetary radius of \rpvaltwo \ \rj, and a mass of \mpvaltwo \ \mj, it is the brightest host among the population of inflated Hot Jupiters ($R_P > 1.5R_J$), making it a valuable discovery for probing the nature of inflated planets. In addition, its existence within a hierarchical triple and its proximity to Earth ($210$ pc) provides a unique opportunity for dynamical studies with continued monitoring with high resolution imaging and precision radial velocities. In particular, the motion of the binary stars around each other and of both stars around the primary star relative to the measured epoch in this work should be detectable when it rises in October 2015.

\end{abstract}

\keywords{exoplanet, transit, hot Jupiter, inflated, hierarchical triple}
\maketitle

\section{Introduction}
\label{sec:intro}

When Hot Jupiters were first discovered \citep{mayor95}, our understanding of planet formation and evolution was turned on its head. However, their existence made transit searches from the ground practical. Although the first transiting planets were originally discovered from follow-up of RV candidates \citep[e.g.][]{charbonneau00, henry00}, the first detections from dedicated transit surveys followed soon after by TrES \citep{alonso04}, XO \citep{mccullough05}, HAT \citep{bakos02}, and WASP \citep{cameron07}, all of which had the same basic design: a small telescope with a wide field of view to monitor many stars to find the few that transited.

The Kilodegree Extremely Little Telescope (KELT) \citep{pepper07} is the most extreme of the mature transit surveys, with the largest single-camera field of view (26 degrees on a side) and the largest platescale (23''/pixel) -- similar to the planned TESS mission \citep{ricker10}. Therefore, while KELT is optimized to find fewer planets, it can find those around brighter host stars which allows a greater breadth and ease of follow-up to fully utilize the wealth of information the transiting planets potentially offer: planetary radius, orbital inclination (and so the true mass), stellar density \citep{seager03}, composition \citep{guillot05, sato05, charbonneau06, fortney06}, spin-orbit misalignment \citep{queloz00, winn05, gaudi07, triaud10}, atmosphere \citep{charbonneau02, vidal03} to name a few -- see \citet{winn10a} for a comprehensive review.

We now describe the discovery of \pname, an inflated Hot Jupiter (R=\rpvaltwo \ \rj) orbiting the bright component (V=10) of a hierarchical triple. In terms of size, \pname \ is qualitatively similar to WASP-79b \citep{smalley12} and WASP-94Ab \citep{neveu14}, which have slightly larger planets around slightly fainter stars. Its size is also similar to KELT-8b \citep{fulton15}. See section 6.2 of \citet{fulton15} for a more detailed comparison of similar planets. 
\pname \ is only the third known transiting planet in a hierarchical triple stellar system, along with WASP-12b and HAT-P-8b \citep{bechter14}. KELT-4 is the brightest host of all these systems, and therefore a valuable find for extensive follow-up of both inflated planets and hierarchical architectures. Because it is relatively nearby (210 pc), continued AO imaging will be able to provide dynamical constraints on the stellar system.

\section{Discovery and Follow-up Observations}
\label{sec:disc}
The procedure we used to identify \pname \ is identical to that described in \citet{siverd12}, using the setup described in \citet{pepper07}, both of which we summarize briefly here.

\pname \ was discovered in field 06 of our survey, which is a $26^\circ$ x $26^\circ$ field of view centered at J2000 09:46:24.1, +31:39:56, best observed in February. We took 150 second exposures with our 42 mm telescope located at Winer Observatory\footnote{\url{http://winer.org/}} in Sonoita, Arizona, with a typical cadence of 15-30 minutes as we cycled between observable fields.

\begin{figure}[!htbp]
  \begin{center}
    \includegraphics[width=3.25in]{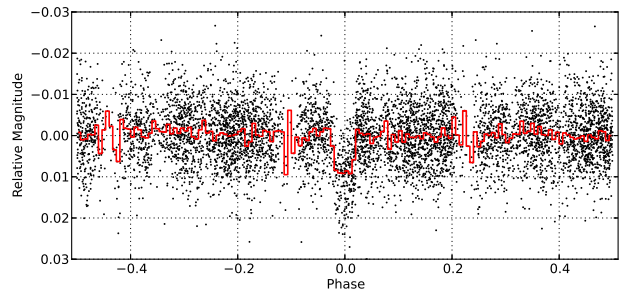} 
    \caption{The TFA-detrended discovery light curve for \pname \ showing all 6571 data points KELT collected from 2006-10-27 to 2011-04-01 (when it was flagged for radial velocity follow-up), phase folded at the best BLS period of 2.9895365 days. The red line shows the data binned in 150 bins ($\sim29$ minutes).}
    \label{fig:discovery}
  \end{center}
\end{figure}

Each object in the KELT survey is matched to the Tycho-2 \citep{hog00} and 2MASS \citep{cutri03,skrutskie06} catalogs, which we use to derive a reduced proper motion cut to remove giants from our sample \citep{cameron07}. After image subtraction, outliers are clipped and the light curves are detrended with the Trend Filtering Algorithm \citep{kovacs05}, and a BLS search is performed \citep{kovacs02}. After passing various programmatic cuts described in \citet{siverd12}, candidates are inspected by eye and selected for follow-up. Figure \ref{fig:discovery} shows the KELT discovery light curve for \pname.

\subsection{SuperWASP}

\begin{figure}[!htbp]
  \begin{center}
    \includegraphics[width=3.25in]{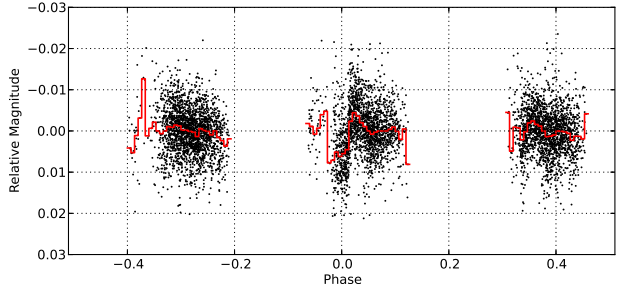} 
    \caption{The observations from two good seasons and cameras of public SuperWASP data for
      \pname \ \citep{butters10}, showing 6020 points from 2006-04-01
      to 2007-05-05, phase folded at the KELT BLS period of 2.9895365
      days. The red line shows the data binned in 150 bins ($\sim29$
      minutes). Despite observing the star at roughly the same
      precision and obtaining nearly as many data points as KELT, the
      near-integer period of \pname \ and relatively short span of
      observations makes SuperWASP's phase coverage sparse, which explains why SuperWASP was not able to identify \pname \ in their data.}
    \label{fig:wasp}
  \end{center}
\end{figure}

As part of the by-eye object selection, we inspect the corresponding public SuperWASP data, if available \citep{butters10}. While SuperWASP achieved roughly the same photometric precision with almost as many observations as KELT did, due to the near-integer period of \pname \ (Period = \pervaltwo) and the relatively short span of the SuperWASP observations, SuperWASP did not observe the ingress of the planet, as shown in Figure \ref{fig:wasp}.

While HAT and SuperWASP adopt a strategy to change fields often,
likely because of the shallow dependence of the detectability with the
duration of observations \citep{beatty08}, KELT has generally opted to
monitor the same fields for much longer, increasing its sensitivity to
longer and near-integer periods, as demonstrated by this find.

\subsection{Follow-up photometry}
\begin{figure}
  \begin{center}
    \includegraphics[height=8.75in,width=0.95\columnwidth]{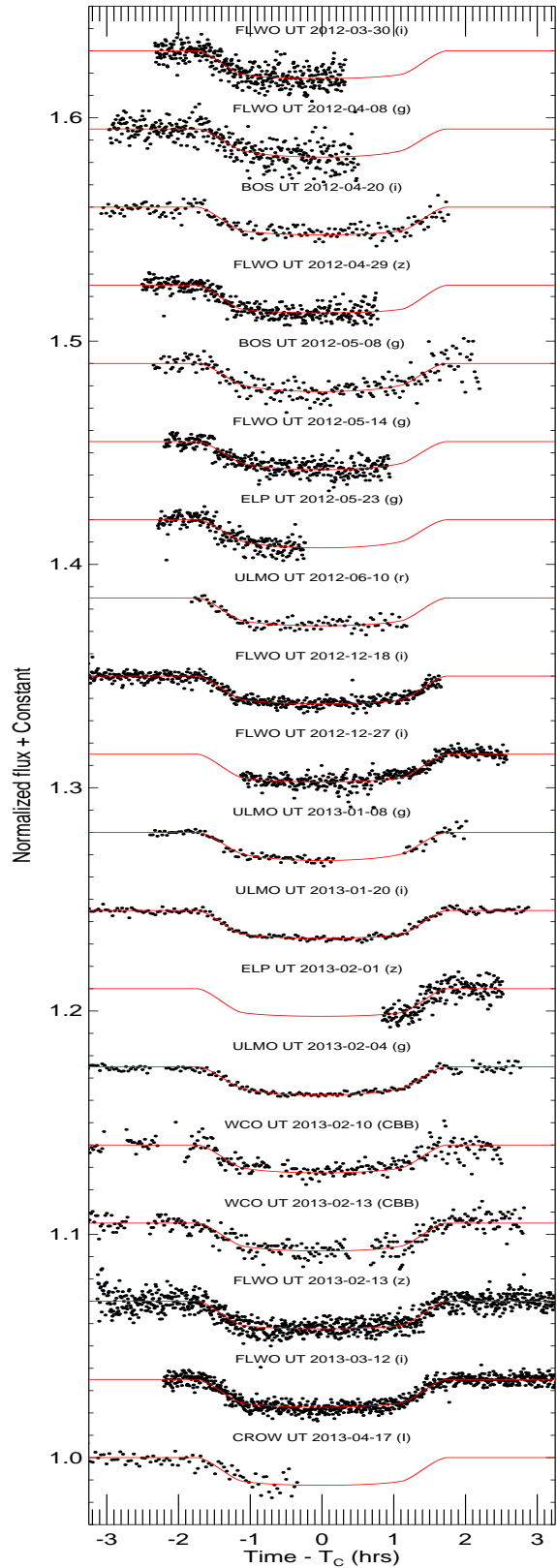} 
    \caption{The 19 follow-up light curves analyzed for \pname, in
      black, with the best-fit model (see \S \ref{sec:model}) shown
      in red. Each light curve has had the SED-modeled contamination from the stellar companion subtracted, the out of transit flux
      normalized to unity and offset by an arbitrary constant
      for clarity, the best fit transit time, $T_C$ subtracted, and a
      trend with airmass removed. The labels above each light curve display the
      telescope, UT date, and filter corresponding to each
      observation.}
    \label{fig:transits}
  \end{center}
\end{figure}

\begin{figure}
  \begin{center}
    \includegraphics[width=3.25in]{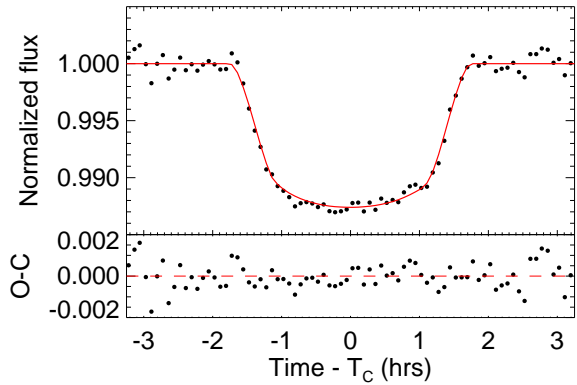} 
    \caption{(Top panel) The 19 follow-up light curves for \pname,
      binned in 5-minute intervals. This is not used during analysis,
      but is just to show the statistical power of the combined fit to
      the entire dataset, as well as the level of systematics present
      in this combined fit. Overlaid is the model for each of the 19
      light curves averaged in the same way. (Bottom panel) The
      residuals of the binned light curve from the binned model in the
      top panel.}
    \label{fig:bintran}
  \end{center}
\end{figure}

We have amassed an extensive follow-up network consisting of around 30
telescopes from amateurs, universities, and professional
observatories. Coordinating with the KELT team, collaborators obtained 19 high-quality transits in six
bands with six different telescopes, shown in Figure
\ref{fig:transits}. All transits are combined and binned in 5-minute
intervals in Figure \ref{fig:bintran} to demonstrate the statistical
power of the combined fit to the entire dataset, as well as the level
of systematics present in this combined fit, though this combined light curve was not used
directly for analysis.

We used KeplerCam on the 1.2 meter Fred Lawrence Whipple Observatory (FLWO) telescope at Mount Hopkins to observe ten transits of \pname \ in the Sloan $i$, $g$, and $z$ bands. These are labeled ``FLWO'' in Figure \ref{fig:transits}. In the end, only eight of these transits were used in the final fit. We excluded one transit on the night of UT 2013-01-15. Cloudy weather forced the dome to close twice during the first half of the observations. Thin clouds continued throughout egress. When analyzed, these data produced a $7\sigma$ significant outlier in the transit time, hinting at large systematics in this lightcurve. We include it with our electronic tables for completeness, but due to the cloudy weather, we do not include it in our analysis. We also excluded another transit observed on the night of UT 2012-05-11. While there is no obvious fault with the light curve, our MCMC analysis found two widely separated, comparably likely regions of parameter space by exploiting a degeneracy in the baseline flux and the airmass detrending parameter, which significantly degraded the quality of the global analysis. Again, we include this observation in the electronic tables for completeness, but do not use it for our analysis. The change in the best-fit parameters of \pname \ is negligible whether this transit is included or not.


We observed four transits, two in the Sloan $g$, one in Sloan $r$, and one
in Sloan $i$, at the Moore Observatory using the 0.6m RCOS telescope,
operated by the University of Louisville in Kentucky (labeled
``ULMO'') and reduced with the AstroImageJ package \citep{collins13b,collins15}.
See \citet{collins14} for additional observatory information.


We observed two transits at the Westminster College Observatory in
Pennsylvania (labeled ``WCO'') with a Celestron C14 telescope in
the CBB (blue blocking) filter. As there are no limb darkening tables
for this filter in \citet{claret11}, we modeled it as the closest
analog available -- the COnvection ROtation and planetary Transits
(CoRoT) bandpass \citep{baglin06}.


Las Cumbres Observatory Global Telescope (LCOGT) consists of a 0.8 meter prototype telescope and nine 1-meter telescopes spread around the world \citep{brown13}. We used the prototype telescope (labeled ``BOS'') to observe transits in the Sloan $i$ band and the Sloan $g$ band. Additionally, we used the 1 meter telescope at McDonald Observatory in Texas (labeled ``ELP'') to observe two partial transits in the Sloan $g$ band and Pan Starrs $z$ band. As there are no limb darkening tables for Pan-Starrs $z$ in \citet{claret11}, we modeled it as the closest analog available -- the Sloan $z$ filter.


We observed one partial transit in the $I$ band at Canela's Robotic Observatory (labeled ``CROW'') in Portugal on UT 2013-04-17. The observations were obtained using a 0.3m LX200 telescope with an SBIG ST-8XME 1530x1020 CCD, giving a 28'x 19' field of view and 1.11 arcseconds per pixel.

\subsection{Radial Velocity}
\label{sec:rv}

\begin{figure}
  \begin{center}
    \includegraphics[width=3.25in]{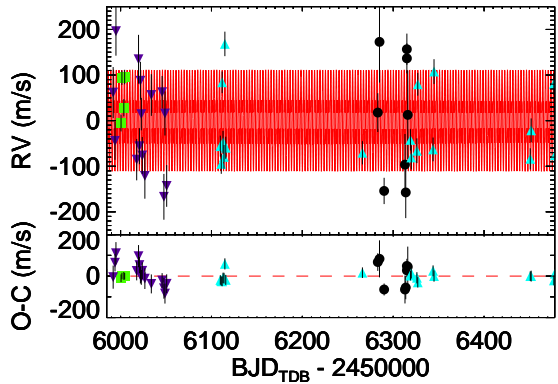} 
    \caption{(Top panel) The unphased RVs for KELT-4A showing HIRES (blue upward triangles), FIES (green squares), EXPERT (black circles), and TRES (purple downward triangles) and the best-fit model in red. The systemic velocity of $-23.5~{\rm km~s^{-1}}$ has been subtracted for clarity. (Bottom panel) The residuals of the RV data from the model fit.}
    \label{fig:rvunphased}
  \end{center}
\end{figure}

\begin{figure}[!htbp]
  \begin{center}
    \includegraphics[width=3.25in]{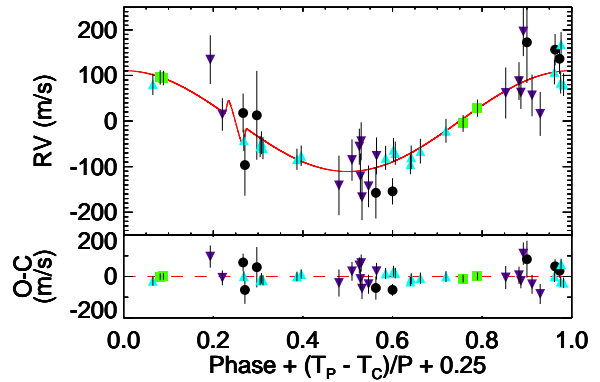} 
    \caption{(Top panel) The phased RV curve for \pname \ showing HIRES (blue upward triangles), FIES (green squares), EXPERT (black circles), and TRES (purple downward triangles). The best-fit model is shown in red, including the Rossiter-McLaughlin (RM) effect. Note that the spin-orbit alignment is extremely poorly constrained by the three points in three different transits and does not conclusively exclude any value. The units of the x-axis were chosen such that the time of transit is centered at 0.25. (Bottom panel) The residuals of the RV data from the model fit.}
    \label{fig:rv}
  \end{center}
\end{figure}

We obtained Radial Velocity (RV) measurements of KELT-4A from three different telescopes/instruments, shown in Figures \ref{fig:rvunphased} and \ref{fig:rv}, and summarized in Table \ref{tab:rv}. The table expresses the radial velocities as relative velocities, using the raw velocities and subtracting the best-fit instrumental velocities from each. For the HIRES velocities, absolute RVs were measured with respect to the telluric lines separately, using the method described by \citet{chubak12} with a mean offset of $-23.5 \pm 0.1~{\rm km~s^{-1}}$.

\begin{figure}[!htbp]
  \begin{center}
    \includegraphics[width=3.25in]{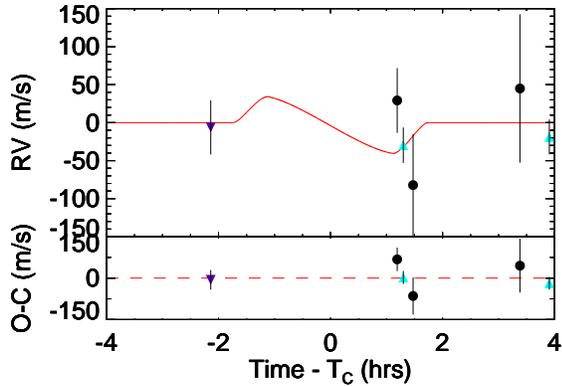} 
    \caption{(Top panel) The phased Rossiter-McLaughlin effect for KELT-4A with the planetary radial velocity signal subtracted showing the one HIRES point and two EXPERT points taken in transit, as well as a few other data points near transit, with the same legend as figures \ref{fig:rvunphased} and \ref{fig:rv}. The best-fit model is shown in red, using the \citet{ohta05} model. Note that the spin-orbit alignment is extremely poorly constrained by the three points in three different transits and does not conclusively exclude any value. (Bottom panel) The residuals of the RV data from the best-fit model.}
    \label{fig:rm}
  \end{center}
\end{figure}

\begin{center}

\begin{threeparttable}
\caption{RV Observations of KELT-4A\label{tab:rv}}
\begin{tabular}{lrrl}
\tableline
\multicolumn{1}{c}{BJD} & \multicolumn{1}{c}{RV\tnote{a}} & \multicolumn{1}{c}{RV error\tnote{b}} & Source \\
\multicolumn{1}{c}{(TDB)} & \multicolumn{1}{c}{(m s$^{-1}$)} & \multicolumn{1}{c}{(m s$^{-1}$)} & \\
\tableline
2455984.708730 &   -144.39 &  31.28 & TRES \\
2455991.800887 &     58.13 &  26.81 & TRES \\
2455993.825798 &    -47.76 &  19.93 & TRES \\
2455994.907665 &    192.71 &  25.75 & TRES \\
2456000.485110 &     -0.50 &  18.50 & FIES \\
2456001.474912 &     99.00 &  15.90 & FIES \\
2456003.570752 &     32.70 &  18.80 & FIES \\
2456004.443338 &    100.20 &  15.90 & FIES \\
2456017.681414 &    -89.66 &  21.74 & TRES \\
2456019.725231 &    130.45 &  25.66 & TRES \\
2456020.719807 &    -60.29 &  18.91 & TRES \\
2456021.785077 &     83.22 &  19.91 & TRES \\
2456022.795489 &     10.00 &  17.07 & TRES \\
2456023.824122 &    -80.68 &  19.33 & TRES \\
2456026.706127 &   -125.61 &  23.94 & TRES \\
2456033.830514 &     51.58 &  21.88 & TRES \\
2456045.714834 &     57.37 &  17.07 & TRES \\
2456047.644014 &   -172.04 &  24.07 & TRES \\
2456048.833086 &     10.57 &  23.13 & TRES \\
2456050.677247 &   -148.27 &  21.92 & TRES \\
2456109.745582 &    -48.13 &   3.36 & HIRES \\
2456110.748739 &    -86.97 &   3.19 & HIRES \\
2456111.750845 &     91.86 &   3.46 & HIRES \\
2456112.744005 &    -39.81 &   3.66 & HIRES \\
2456113.743451 &    -71.90 &   3.80 & HIRES \\
2456114.743430 &    176.00 &   3.92 & HIRES \\
2456115.743838 &    -51.79 &   3.54 & HIRES \\
2456266.106636 &    -67.17 &   3.72 & HIRES \\
2456283.028749 &     18.00 &  31.00 & EXPERT\tablenotemark{c} \\
2456284.922216 &    173.00 &  65.00 & EXPERT \\
2456290.004820 &   -154.00 &  21.00 & EXPERT \\
2456312.936489 &    -97.00 &  49.00 & EXPERT\tablenotemark{c} \\
2456313.810429 &   -158.00 &  41.00 & EXPERT \\
2456315.006329 &    156.00 &  25.00 & EXPERT \\
2456315.035867 &    136.00 &  24.00 & EXPERT \\
2456316.005608 &     12.00 &  71.00 & EXPERT \\
2456318.908411 &    -40.54 &   3.41 & HIRES\tablenotemark{c} \\
2456319.854373 &    -79.04 &   3.83 & HIRES \\
2456326.065752 &    -64.53 &   3.96 & HIRES \\
2456327.021240 &     81.25 &   3.64 & HIRES \\
2456343.823502 &    -62.11 &   3.80 & HIRES \\
2456344.897907 &    108.59 &   3.96 & HIRES \\
2456450.804949 &    -85.90 &   3.39 & HIRES \\
2456451.799327 &    -23.46 &   3.83 & HIRES \\
2456476.749960 &     77.81 &   3.46 & HIRES \\
2456477.739810 &    -79.00 &   3.29 & HIRES \\
\tableline
\end{tabular}

\begin{tablenotes}
    \item[a]{The offsets for each telescope have been fitted and subtracted. The systemic velocity, measured from Keck as $-23.5 \pm 0.1~{\rm km~s^{-1}}$ may be added to each observation to get the absolute velocities.}
    \item[b]{Unscaled measurement uncertainties.}
    \item[c]{Observation occurred during transit and was affected by the Rossiter-McLaughlin effect.}
\end{tablenotes}

\end{threeparttable}
\end{center}

Using the High Resolution Echelle Spectrometer (HIRES) instrument \citep{vogt94} on the Keck I telescope located on Mauna Kea, Hawaii, we obtained 16 exposures between 2012-07-01 and 2013-02-21 with an iodine cell, plus a single iodine-free template spectrum. One of these points fell within the transit window and therefore provides a weak constraint on the Rossiter-McLaughlin effect (see Figure \ref{fig:rm}). We followed standard procedures of the California Planet Survey (CPS) to set up and use HIRES, reduce the spectra, and compute relative RVs \citep{howard10b}.  We used the ``C2'' decker (0.86$''$ wide) and oriented the slit with an image rotator to avoid contamination from KELT-4B,C.

We obtained five spectra with the FIbre-fed Echelle Spectrograph (FIES) on the 2.5 meter Nordic Optical Telescope (NOT) in La Palma, Spain \citep{djupvik10} between 2012-03-13 and 2012-03-17 with the high-resolution fiber ($1\farcs 3$ projected diameter) with resolving power $R \approx 67,000$. We discarded one observation which the observer marked as bad and had large quoted uncertainites. We used standard procedures to reduce these data, as described in \citet{buchhave10} and \citet{buchhave12}.

Eight spectra were taken with the EXPERT spectrograph \citep{ge10} at the 2.1m telescope at Kitt Peak National Observatory between 2012-12-21 and 2013-01-23 and reduced using a modified pipeline described by \citet{wang12}. EXPERT has a resolution of R=30,000, 0.39-1.0 $\mu$m coverage, and a $1\farcs 2$ fiber. Two of these spectra were taken during transit and therefore provide a weak constraint on the RM effect.

We took 16 radial velocity observations with the TRES spectrograph \citep{furesz08}, which has a resolving power of 44,000, a fiber diameter of $2\farcs 3$, and a typical seeing of $1 \farcs 5$. Because of the typical seeing, the fiber diameter, and the nearby companion, we initially excluded all of the TRES data, but when we found the fit to be consistent (albeit with slightly higher scatter than is typical for TRES), we included it in the global fit. The higher scatter was taken into account by scaling the errors such that the probability of \chisq \ we got was 0.5, as we do for all data sets.

Stellar parameters (\logg, \teff, \feh, and \vsini) were derived from the FIES and TRES using the Stellar Parameter Classification (SPC) tool \citep{buchhave14}, and the HIRES spectra using SpecMatch (Petigura et al. 2015 in prep.). It is well known that the transit lightcurve alone can constrain the stellar density \citep{seager03}. Coupled with the YY isochrones and a measured \teff, the transit light curve provides a tight constraint on the \logg \ \citep{torres12}. Alternatively, we have discovered that the limb darkening of the transit itself is sufficient to loosely constrain the \teff \ and therefore the \logg \ without spectroscopy during a global fit, which we use as another check on the stellar parameters. Finally, we iterated on the HIRES spectroscopic parameters using a prior on the \logg \ from the global fit. That is, we used the HIRES spectroscopic parameters to seed a global fit, found the \logg \ using the more precise transit constraint, fed that back into SpecMatch to derive new stellar parameters that are consistent with the transit, and then ran the final global fit.

All of these methods were marginally consistent ($< 1.7\sigma$) with one another, as shown in Table \ref{tab:specpars}. Since the uncertainties in the measured stellar parameters are typically dominated by the stellar atmospheric models, this marginal consistency is uncommon and may be indicative of a larger than usual systematic error. It is likely that the discrepancy is due to the blend with the neighbor $1\farcs 5$ away. While the FIES \logg \ agrees best with the \logg \ derived from the transit photometry, we adopted the HIRES parameters derived with an iterative \logg \ prior from the global analysis because of its higher spatial resolution and better median site seeing. However, to account for the inconsistency between methods, we inflated the uncertainties in \teff \ and \feh \ as shown in Table \ref{tab:specpars} so they were in good agreement with the values without the \logg \ prior and did not include a spectroscopic prior on the \logg \ during the global fit. Still, systematic errors in the stellar parameters (and therefore the derived planetary parameters) at the 1-sigma level would not be surprising. The slightly hotter star preferred by the other spectroscopic methods would make the star bigger and therefore the planet even more inflated. The cooler star preferred by the limb darkening would make the star and planet smaller.

\begin{deluxetable*}{lcccc}
\tablecaption{Summary of measured stellar parameters}
\tablehead{
\colhead{} & \colhead{\logg} & \colhead{\teff} & \colhead{\feh} & \colhead{\vsini} \\
\colhead{Instrument} & \colhead{cgs} & \colhead{K} & \colhead{} & \colhead{${\rm km~s^{-1}}$}
}
\startdata
FIES                          & $4.11  \pm 0.10$  & $6360 \pm 49$        & $-0.12 \pm 0.08$ & $7.6 \pm 0.5$\\
TRES                          & $4.05  \pm 0.10$  & $6249 \pm 49$        & $-0.12 \pm 0.08$ & $7.8 \pm 0.5$\\
HIRES                         & $4.20  \pm 0.08$  & $6281 \pm 70$        & $-0.10 \pm 0.05$ & $7.6 \pm 1.7$\\
HIRES\tablenotemark{a}        & $4.12  \pm 0.08$  & $6218 \pm 70$        & $-0.12 \pm 0.05$ & $6.2 \pm 1.2$\\
Global fit\tablenotemark{b}   & $4.104 \pm 0.019$ & $6090^{+390}_{-320}$ & --               & --           \\
Adopted priors\tablenotemark{c} & N/A           & $6218 \pm 80$        & $-0.12 \pm 0.08$   & $6.2 \pm 1.2$\\
Final values\tablenotemark{d} & \loggstarvaltwo   & \teffvaltwo          & \fehvaltwo       & $6.2 \pm 1.2$\\
\enddata
\tablenotetext{a}{Includes an iterative \logg \ prior from the global transit fit.}
\tablenotetext{b}{Values from the global fit without a \logg \ or \teff \ prior, but with an $\feh=-0.12 \pm 0.08$ prior and guided by the stellar limb darkening.}
\tablenotetext{c}{Spectroscopic priors used in the final iteration of the global fit.}
\tablenotetext{d}{The values from the final iteration of the global fit with the adopted spectroscopic priors.}
\label{tab:specpars}
\end{deluxetable*}

\subsection{Historical Data}
\label{sec:history}

As compiled by the Washington Double Star Catalog \citep{mason01},
KELT-4 was originally identified as a common proper motion binary with a separation of 1.5'' by \citet{couteau73}, who named it COU 777. It was later observed in 1987 by \citet{argue92}, {\it Hipparcos} in 1991
\citep{perryman97,vanleeuwen07}, and the Tycho Survey in 1991
\citep{fabricius02}. The magnitudes, position angles (degrees East of
North), and separations (arcseconds) from these historical records are
summarized in Table \ref{tab:binary} at the observed epochs, in
addition to our own measurement described in \S \ref{sec:ao}.

\begin{deluxetable*}{ccccccccccccccl}[!htbp]
\tablecaption{The positions of the components of KELT-4 from historical data.}
\setlength{\tabcolsep}{0.02in} 
\tabletypesize{\scriptsize}
\tablehead{
  \colhead{} & 
  \colhead{$PA_{A,BC}$} & 
  \colhead{$Sep_{A,BC}$} & 
  \colhead{$PA_{BC}$} & 
  \colhead{$Sep_{BC}$} &
  \colhead{}& 
  \colhead{}   & 
  \colhead{} & 
  \colhead{}   & 
  \colhead{} & 
  \colhead{}  &  
  \colhead{} & 
  \colhead{}  & 
  \colhead{}  \\
  \colhead{Epoch} &
  \colhead{(degrees)} &
  \colhead{(arcsec)} &
  \colhead{(degrees)} & 
  \colhead{(mas)} & 
  \colhead{$V_{A}$} & 
  \colhead{$V_{BC}$} & 
  \colhead{$R_{A}$} & 
  \colhead{$R_{BC}$} & 
  \colhead{$J_{A}$} & 
  \colhead{$J_{BC}$} & 
  \colhead{$K_{A}$} & 
  \colhead{$K_{BC}$} & 
  \colhead{Source}
}
\startdata
1972.230 & 38.9 & 1.430                   & --            & --           &  9.500 & 14.000 &    -- &      -- &     -- &     -- &    -- &     -- & 1 \\
1972.249 & 34.0 & 1.570                   & --            & --           &  9.500 & 14.000 &    -- &      -- &     -- &     -- &    -- &     -- & 1 \\
1987.430 & 35.0 & 1.380                   & --            & --           & 10.130 & 12.370 &  9.81 &   11.96 &     -- &     -- &    -- &     -- & 2 \\
1991.250 & 31.0 & $1.553 (43)$            & --            & --           & 10.186 & 12.992 &    -- &      -- &     -- &     -- &    -- &     -- & 3 \\
1991.530 & 33.1 & 1.560                   & --            & --           & 10.042 & 12.992 &    -- &      -- &     -- &     -- &    -- &     -- & 4 \\
2012.3464 & $28.887 (70)$ & $1.5732 (18)$ & $325.23 (13)$ & $49.14 (39)$ & --     &     -- &    -- &      -- &  9.193 &  10.94 & 8.972 &  10.35 & 5
\enddata
\label{tab:binary}
\tablerefs{
1=\citet{couteau73};
2=\citet{argue92};
3=\citet{perryman97,vanleeuwen07}; 
4=\citet{fabricius02};
5=This work}
\end{deluxetable*}

\subsection{High-Resolution Imaging}
\label{sec:ao}

\begin{figure}[!htbp]
  \begin{center}
    \includegraphics[width=3.25in]{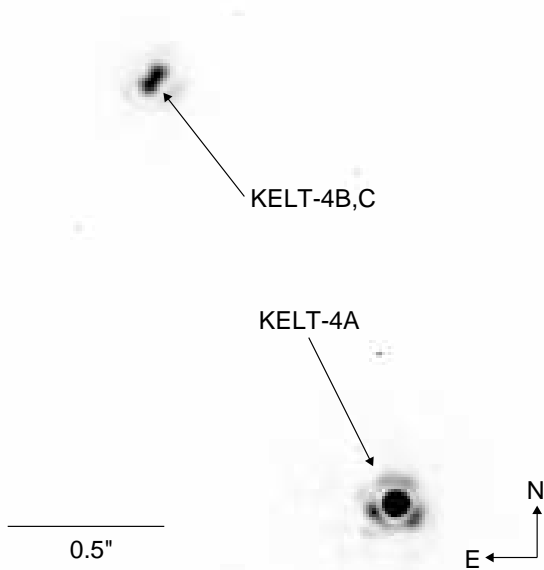} 
    \caption{The KECK AO image taken by the NIRC2 instrument in the K
      band shows KELT-4A in the bottom right, and a physically-bound,
      blended binary star (KELT-4B,C) 1.5'' to the northeast.}
    \label{fig:ao}
  \end{center}
\end{figure}

On 2012-05-07, we obtained adaptive optics (AO) imaging on the Keck II
telescope located on Mauna Kea, Hawaii, using NIRC2 in both the J and
K bands \citep{yelda10}, shown in Figure \ref{fig:ao}. We used the
narrow camera, with a pixel scale of 0.009942''/pix.

The proper motion determined by {\it Hipparcos} of $\mu_{\alpha}=11.79 \pm 1.31$
mas yr$^{-1}$ and $\mu_{\delta}= -12.63 \pm 0.9$ mas yr$^{-1}$ over the 40-year baseline
between the original observations by \citet{couteau73} and ours have
amounted to over 0.5'' of total motion. If the companion mentioned in
\S \ref{sec:history} was not gravitationally bound, this motion would
have significantly changed the separation, which would be trivial to
detect in our AO images. However, the separations remain nearly
identical. Therefore, we confirm this system as a common proper motion
binary.

Interestingly, for the first time, our AO image further resolves the stellar companion as a binary itself, with a separation of $49.14 \pm 0.39$ mas and a position angle of $325.23 \pm 0.13$ degrees at epoch 2012.3464, as shown in Figure \ref{fig:ao}. From their relative magnitudes and SED modeling, we estimate this pair to be twin K stars with $\teff = 4300$ K and $\rs = 0.6 \pm 0.1 \rsun$. Using \citet{demory09}, we translate that to a mass of $0.65 \pm 0.1 \msun$. Therefore, \pname \ is a companion to
the brightest member of a hierarchical triple stellar system, similar
to WASP-12b and HAT-P-8b \citep{bechter14}. That is, KELT-4A is
orbited by \pname, a $\sim 1 M_J$ mass planet with a period of 3
days and also by KELT-4BC, a twin K-star binary. In all of our
follow-up light curves, this double, with a combined V magnitude of
13, was blended with KELT-4, contributing $\sim2-7\%$ to the baseline
flux, depending on the observed bandpass.

At the distance of 210 pc determined from our SED modeling (\S \ref{sec:sed}), the projected separation between KELT-4A and KELT-4BC is $328 \pm 16$ AU, and the projected separation KELT-4B and KELT-4C is $10.3 \pm 0.74$ AU. Assuming the orbit is face on and circular, the period of the outer binary, $P_{A,BC}$, would be $3780 \pm 290$ years and the period of the twin stars, $P_{B,C}$, would be $29.4 \pm 3.6$ years. 

While assuming the orbit is face on and circular is likely incorrect, it gives us a rough order of magnitude of the signal we might expect. If correct, in the 21 years between {\it Hipparcos} and Keck data, we would expect to see about 55 mas of motion of KELT-4BC relative to KELT-4A. This is roughly what we see, though it is worth noting that the clockwise trend between the {\it Hipparcos} position and our measurement is in contradiction to the 1-sigma counter-clockwise trend for the two {\it Hipparcos} values.

\begin{figure}
  \begin{center}
    \includegraphics{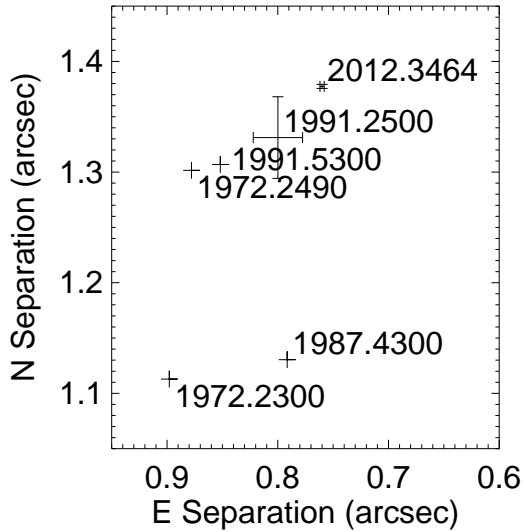} 
    \caption{The positions of KELT-4BC relative to KELT-4A from Table \ref{tab:binary}. The epochs of each observation are printed next to the corresponding data point, showing the slow clockwise motion. Note that the data point from this work has error bars smaller than the point, and several points do not have quoted uncertainties.}
    \label{fig:pos}
  \end{center}
\end{figure}

Extending the baseline to the full 40 years, we expect to see about 110 mas of motion. While the historical values are quoted without uncertainties, using the 200 mas discrepancy between the two 1972 data points as a guide, the data seem to confirm the clockwise trend and are consistent with a 110 mas magnitude (see Figure \ref{fig:pos}). Unfortunately, the data sample far too little of the orbit and are far too imprecise to provide meaningful constraints on any other orbital parameters.

\section{Modeling}
\label{sec:model}

\subsection{SED}
\label{sec:sed}

We used the broadband photometry for the combined light for all three components, summarized in Table
\ref{tab:starprops}, the spectroscopic value of \teff, and an
iterative solution for \rs \ from EXOFAST to model the SED
of KELT-4A and KELT-4BC, assuming the B and C components were
twins.

\begin{deluxetable*}{llclc}
\centering
\tablecolumns{5}
\tablewidth{0pt}
\tabletypesize{\scriptsize}
\tablecaption{Stellar Properties of KELT-4A and combined photometry for all three components used for the SED fit.\label{tab:starprops}}
\tablehead{\colhead{~~~Parameter} & \colhead{Description (Units)} & \colhead{Value}  & \colhead{Source} & \colhead{Reference}  }
\startdata
Names              &                                      & BD+26 2091                 &           &   \\
                   &                                      & HIP 51260                  &           &   \\
                   &                                      & GSC 01973-00954            &           &   \\
                   &                                      & SAO 81366                  &           &   \\
                   &                                      & 2MASS J10281500+2534236    &           &   \\
                   &                                      & TYC 1973 954 1             &           &   \\
                   &                                      & CCDM J10283+2534A          &           &   \\
                   &                                      & WDS 10283+2534             &           &   \\
                   &                                      & GALEX J102814.9+253423     &           &   \\ 
                   &                                      & COU 777                    &           &   \\
$\alpha_{J2000}$   & Right Ascension (J2000 )             & 10 28 15.011               & Hipparcos & 1 \\
$\delta_{J2000}$   & Declination (J2000)                  & +25 34 23.47               & Hipparcos & 1 \\

FUV$_{GALEX}$      & Far UV Magnitude                     & $20.39 \pm 0.16$           &  GALEX    & 2 \\
NUV$_{GALEX}$      & Near UV Magnitude                    & $14.49 \pm 0.01$           &  GALEX    & 2 \\

B                  & Johnson B Magnitude                  & $10.47 \pm 0.03$           & APASS     & 3 \\
V                  & Johnson V Magnitude                  & $9.98 \pm 0.03$            & APASS     & 3 \\

%
%

$J$                & J Magnitude                          & $9.017 \pm 0.021$          & 2MASS     & 4 \\
$H$                & H Magnitude                          & $8.790 \pm 0.023$          & 2MASS     & 4 \\
$K$                & K Magnitude                          & $8.689 \pm 0.020$          & 2MASS     & 4 \\


WISE1              & WISE 3.6 $\mu$m                      & $8.593 \pm 0.022$          & WISE      & 5 \\
WISE2              & WISE 4.6 $\mu$m                      & $8.642 \pm 0.020$          & WISE      & 5 \\     
WISE3              & WISE 11 $\mu$m                       & $8.661 \pm 0.023$          & WISE      & 5 \\  
WISE4              & WISE 22 $\mu$m                       & $8.608 \pm 0.336$          & WISE      & 5 \\
$\mu_{\alpha}$     & Proper Motion in RA (mas yr$^{-1}$)  & $11.79 \pm 1.31$           & Hipparcos & 1 \\  
$\mu_{\delta}$     & Proper Motion in Dec (mas yr$^{-1}$) & $-12.63 \pm 0.90$           & Hipparcos & 1 \\  
$\pi$\tablenotemark{a}           & Parallax (mas)                       & $3.02 \pm 1.41$            & Hipparcos & 1 \\
$d$\tablenotemark{a}                 & Distance (pc)                      & \dvalone                & This work (Eccentric)&     \\ 
$d$\tablenotemark{a}                 & Distance (pc)                      & \dvaltwo                & This work (Circular)&     \\  
$d$\tablenotemark{a}                 & Distance (pc)                      & $ 210 \pm 10$                & This work (SED)&     \\ 
$U$\tablenotemark{b} & Galactic motion (${\rm km~s^{-1}}$) & $33.2 \pm 1.3$              & This work &     \\  
$V$                  & Galactic motion (${\rm km~s^{-1}}$) & $9.8 \pm 1.0$              & This work &     \\  
$W$                  & Galactic motion (${\rm km~s^{-1}}$) & $-8.6 \pm 0.7$              & This work &     \\  
$A_V$                & Visual Extinction                  & $0.05^{+0.0}_{-0.03}$           & This work &     \\  
\enddata
\tablenotetext{a}{We quote the parallax from Hipparcos, but derive a more precise distance from our SED modeling and semi-independently through both the eccentric and circular global EXOFAST models. While all are consistent, we adopt the SED distance as our preferred value.}
\tablenotetext{b}{Positive $U$ is in the direction of the Galactic Center.}
\tablerefs{
1=\citet{perryman97,vanleeuwen07}; 
2=\citet{martin05b};               
3=\citet{henden12};                
4=\citet{cutri03,skrutskie06};     
5=\citet{wright10, cutri12};       
}
\end{deluxetable*}

\begin{figure}
  \begin{center}
    \includegraphics[width=2.5in,angle=90]{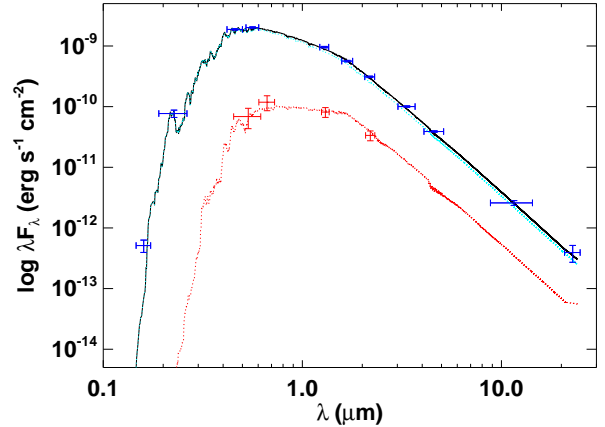} 
    \caption{Measured and best-fit SEDs for KELT-4A (cyan), the combined SED from KELT-4B and KELT-4C (red) and the combined light from KELT-4A, KELT-4B, and KELT-4C (black) from UV through NIR.  The error bars indicate measurements of the flux in UV, optical, and NIR passbands listed in Tables \ref{tab:starprops} and \ref{tab:binary}. The vertical errorbars are the $1\sigma$ photometric uncertainties, whereas the horizontal error bars are the effective widths of the passbands. The solid curves are the best-fit theoretical SED from the NextGen models of \citet{hauschildt99}, assuming stellar parameters \teff, \logg, and \feh \ fixed at the values in Table  \ref{tab:starprops} from the circular fit, with $A_V$ and d allowed to vary.}
    \label{fig:sed}
  \end{center}
\end{figure}

The SED of KELT-4 is shown in Figure \ref{fig:sed}, using the blended photometry from all 3 components summarized in Table \ref{tab:starprops}. We fit for extinction, $A_V$ and the distance, $d$. $A_V$ was limited to a maximum of 0.05 based on the Schlegel dust map value \citep{schlegel98} for the full extinction through the Galaxy along the line of sight). From the SED analysis, we derive an extinction of 0.01 mag and a distance of $210 \pm 10$ pc -- consistent with, but much more precise than the {\it Hipparcos} parallax ($330 \pm 150$ pc).

Because KELT-4A was blended with KELT-4BC in all of our transit photometry, the SED-modeled contributions from KELT-4B and KELT-4C were subtracted before modeling the transit. The contribution from the KELT-4BC component was $1.827\%$ in Sloan $g$, $4.077\%$ in Sloan $r$, $5.569\%$ in Sloan $i$, $7.017\%$ in Sloan $z$, and $3.670\%$ in $CBB$. 

We ran several iterations of the EXOFAST fit (see \S \ref{sec:exofast}), first with a prior on the distance from the SED modeling, but without priors on \teff \ or \logg. The \rs \ from EXOFAST was fed back into the SED model, and we iterated until both methods produced consistent values for \teff \ and \logg.

Because the distance derived from the SED modeling relies on the \rs \ from EXOFAST, we removed the distance prior during the final iteration of the global fit so as not to double count the constraint. The only part of the SED modeling our global fit relies on is the extinction and the blending fractions in each bandpass that dilute the transit depth.  The details of the iterative fit are described further in \S \ref{sec:exofast}.

\subsection{Galactic Model}
\label{sec:galmodel}
Using the distance, proper motion, the systemic velocity from Keck, and the \citet{coskunoglu11} determination of the Sun's peculiar motion with respect to the local standard of rest, we calculate the 3-space motion of the KELT-4A system through the Galaxy, summarized in table \ref{tab:starprops}. According to the classification scheme of \citet{bensby03}, this gives the system a 99\% likelihood of being in the Galaxy's thin disk, which is consistent with the other known parameters of the system.

\subsection{Global Model}
\label{sec:exofast}

Similar to \citet{beatty12}, after iterative SED modeling and transit modeling with the blend subtracted converged on the same stellar properties, we used a modified version of EXOFAST \citep{eastman13} to model the unblended KELT-4A parameters, radial velocities, and deblended transits in a global solution.

We imposed Gaussian priors for $\teff = 6218 \pm 80$ K, $\feh=-0.12 \pm 0.08$, and $V\sin{I_*}=6.2\pm1.2~{\rm km~s^{-1}}$ from the Keck high resolution spectra as measured by SpecMatch with an iterative solution on \logg \ from the global fit. The uncertainties in \feh \ and \teff \ were inflated due to the marginal disagreement with the parameters measured by FIES and TRES, as discussed in \S \ref{sec:rv}. 

We also imposed Gaussian priors from a linear fit to the transit times: $P=2.9895933 \pm 0.0000049$ and $T_{C}=2456190.30201 \pm 0.00022$. These priors do not affect the measured transit times since a separate TTV was fit to each transit without limit. These priors only impact the RV fit, the timing of the RM effect, and the shape of the transit slightly through the period. In addition, we fixed the extinction, $A_V$ to 0.01 and the deblending fractions for each band summarized in \S \ref{sec:sed} from the SED analysis, and the V-band magnitude to 10.042 from Tycho in order to derive the distance. 

The errors for each data set were scaled such that the probability of obtaining the \chisq \ we got from an independent fit was 0.5. For all the transit data, the scaled errors are reported in the online data sets. For the eccentric fit to the EXPERT data, which did not have enough data points for an independent fit, we iteratively found the residuals from the global fit and scaled the uncertainties based on that. The FIES data, with only four good data points, also did not have enough data points for independent fits for either the eccentric or circular fits. However, it had a scatter about the best-fit global model that was smaller than expected. We opted not to scale the FIES uncertainties at all, as enforcing a $\rchisq=1$ would result in uncertainties that were significantly smaller than HIRES, which is not justified based on our experience with both instruments. The scalings for each fit and RV data set are reported in Table \ref{tab:exofastpars}. Note that a common jitter term for all data sets does not reproduce a $\rchisq=1$ for each data set, as one would expect if the stellar jitter were the sole cause of the additional scatter. We would require a $55~{\rm m~s^{-1}}$ jitter term for the EXPERT and TRES data sets and a $23~{\rm m~s^{-1}}$ jitter term for the HIRES RVs. The FIES $\rchisq$ is below 1, so no jitter term could compensate. We suppose that contamination is to blame for the higher scatter in the TRES and EXPERT data, while the limited number of data points makes it relatively likely to get a smaller-than-expected scatter by chance for the FIES data.

We replaced the Torres relation within EXOFAST with Yonsie Yale (YY) evolutionary tracks \citep{yi01,demarque04} to derive the stellar properties more consistently with the SED analysis. At each step in the Markov chain, $R_*$ was derived from the step parameters. That, along with the steps in $\log{M_*}$ and $\feh$ were used as inputs to the YY evolutionary tracks to derive a value for \teff. Since there are sometimes more than one value of \teff \ for given values of $\log{M_*}$ and $\feh$, we use the YY \teff \ closest to the step value for \teff. The global model is penalized by the difference between the YY-derived \teff \ and the MCMC step value for \teff, assuming a YY model uncertainty of 50 K, effectively imposing a prior that the host star lie along the YY evolutionary tracks. The step in \teff \ was further penalized by the difference between it and spectroscopic prior in \teff \ to impose the spectroscopic constraint. This same method was used in all KELT discoveries including and after KELT-6b, as well as HD 97658b \citep{dragomir13}.

The distance derived in Table \ref{tab:exofastpars} does not come from an explicit prior from the SED analysis. Rather, the value quoted in the table is derived through the transit \logg \ and the \feh \ and \teff \ priors coupled with the YY evolutionary tracks (i.e., the stellar luminosity), the extinction, the magnitude, and the bolometric correction from \citet{flower96} (and $M_{bol,\Sun}=4.732$, \citet{torres10}). The agreement in the distances derived from EXOFAST and the SED analysis is therefore a confirmation that the two analyses were done self-consistently. We adopt the distance determination from the SED fit ($210 \pm 10$ pc) as the preferred value.

The quadratic limb darkening parameters, summarized in Table \ref{tab:ldpars}, were derived by interpolating the \citet{claret11} tables with each new step in \logg, \teff, and \feh \ and not explicitly fit. Since this method ignores the systematic model uncertainty in the limb darkening tables, which likely dominate the true uncertainty, we do not quote the MCMC uncertainties. All light curves observed in the same filter used the same limb darkening parameters. 

\begin{deluxetable*}{lccc}
\tablecaption{Median values and 68\% confidence interval for the limb darkening parameters for KELT-4A}
\tablehead{\colhead{~~~Parameter} & \colhead{Units} & \colhead{Eccentric} & \colhead{Circular}}
\startdata
$u_{1,CoRoT}$\dotfill &    Linear Limb-darkening\dotfill &      \uonecorotvalone &      \uonecorotvaltwo\\
$u_{2,CoRoT}$\dotfill & Quadratic Limb-darkening\dotfill &      \utwocorotvalone &      \utwocorotvaltwo\\
$u_{1,g'}$\dotfill    &    Linear Limb-darkening\dotfill &     \uonesloangvalone &     \uonesloangvaltwo\\
$u_{2,g'}$\dotfill    & Quadratic Limb-darkening\dotfill &     \utwosloangvalone &     \utwosloangvaltwo\\
$u_{1,r'}$\dotfill    &    Linear Limb-darkening\dotfill &     \uonesloanrvalone &     \uonesloanrvaltwo\\
$u_{2,r'}$\dotfill    & Quadratic Limb-darkening\dotfill &     \utwosloanrvalone &     \utwosloanrvaltwo\\
$u_{1,i'}$\dotfill    &    Linear Limb-darkening\dotfill &     \uonesloanivalone &     \uonesloanivaltwo\\
$u_{2,i'}$\dotfill    & Quadratic Limb-darkening\dotfill &     \utwosloanivalone &     \utwosloanivaltwo\\
$u_{1,I}$\dotfill     &    Linear Limb-darkening\dotfill &          \uoneivalone &          \uoneivaltwo\\
$u_{2,I}$\dotfill     & Quadratic Limb-darkening\dotfill &          \utwoivalone &          \utwoivaltwo\\
$u_{1,z'}$\dotfill    &    Linear Limb-darkening\dotfill &     \uonesloanzvalone &     \uonesloanzvaltwo\\
$u_{2,z'}$\dotfill    & Quadratic Limb-darkening\dotfill &     \utwosloanzvalone &     \utwosloanzvaltwo\\
\enddata
\label{tab:ldpars}
\end{deluxetable*}

We modeled the system allowing a non-zero eccentricity of \pname, but found it perfectly consistent with a circular orbit. This is generally expected because the tidal circularization timescales of such Hot Jupiters are much much smaller than the age of the system \citep{adams06}. Therefore, we reran the analysis fixing \pname's eccentricity to zero. The results of both the circular and eccentric global analyses are summarized in Table \ref{tab:exofastpars}, though we generally favor the circular fit due to our expectation that the planet is tidally circularized and the smaller uncertainties. All figures and numbers shown outside of this table are derived from the circular fit.

While we only had 3 serendipitous radial velocity data points during transit, we allowed $\lambda$, the spin-orbit alignment, to be free during the fit. The most likely model is plotted in Figure \ref{fig:rv} and a zoom in on the RM effect in Figure \ref{fig:rm}, showing the data slightly favor an aligned geometry. However, the median value and 68\% confidence interval ($\lambda = \lambdavaltwo$) show this constraint is extremely weak. In reality, the posterior for $\lambda$ is bimodal with peaks at $-30^\circ$ and $120^\circ$, has a non-negligible probability everywhere, and is strongly influenced by the \vsini \ prior. In fact, the distribution of likely values is not far from uniform which would have a 68\% confidence interval of $0 \pm 123$ degrees. Therefore, we consider $\lambda$ to be essentially unconstrained. Note that in our quoted (median) values for angles, we first center the distribution about the mode to prevent boundary effects from skewing the inferred value to the middle of the arbitrary range.

Finally, we fit a separate transit time, baseline flux, and detrend with airmass to each of the 19 transits during the global fit, a separate zero point for each of the 4 RV data sets, and a slope to detect an RV trend, for a total of 75 free parameters (73 for the circular fit).

\begin{deluxetable*}{lccc}
\tablecaption{Median values and 68\% confidence interval for \pname}
\tablehead{\colhead{~~~Parameter} & \colhead{Units} & \colhead{Eccentric} & \colhead{Circular}}
\startdata
\sidehead{Stellar Parameters:}
            ~~~$M_{*}$\dotfill &                              Mass (\msun)\dotfill &          \mstarvalone &          \mstarvaltwo\\
            ~~~$R_{*}$\dotfill &                            Radius (\rsun)\dotfill &          \rstarvalone &          \rstarvaltwo\\
            ~~~$L_{*}$\dotfill &                        Luminosity (\lsun)\dotfill &          \lstarvalone &          \lstarvaltwo\\
           ~~~$\rho_*$\dotfill &                             Density (cgs)\dotfill &            \rhovalone &            \rhovaltwo\\
              ~~~$Age$\dotfill &                                 Age (Gyr)\dotfill &            \agevalone &            \agevaltwo\\
        ~~~$\log{g_*}$\dotfill &                     Surface gravity (cgs)\dotfill &       \loggstarvalone &       \loggstarvaltwo\\
            ~~~$\teff$\dotfill &                 Effective temperature (K)\dotfill &           \teffvalone &           \teffvaltwo\\
             ~~~$\feh$\dotfill &                               Metallicity\dotfill &            \fehvalone &            \fehvaltwo\\
       ~~~$v\sin{I_*}$\dotfill &                 Rotational velocity (m/s)\dotfill &      \vsinistarvalone &      \vsinistarvaltwo\\
          ~~~$\lambda$\dotfill &            Spin-orbit alignment (degrees)\dotfill &         \lambdavalone &         \lambdavaltwo\\
                ~~~$d$\dotfill &                             Distance (pc)\dotfill &              \dvalone &              \dvaltwo\\
\sidehead{Planetary Parameters:}
                ~~~$e$\dotfill &                              Eccentricity\dotfill &              \evalone &              \evaltwo\\
         ~~~$\omega_*$\dotfill &          Argument of periastron (degrees)\dotfill &          \omegavalone &          \omegavaltwo\\
                ~~~$P$\dotfill &                             Period (days)\dotfill &              \pvalone &              \pvaltwo\\
                ~~~$a$\dotfill &                      Semi-major axis (AU)\dotfill &              \avalone &              \avaltwo\\
            ~~~$M_{P}$\dotfill &                                Mass (\mj)\dotfill &             \mpvalone &             \mpvaltwo\\
            ~~~$R_{P}$\dotfill &                              Radius (\rj)\dotfill &             \rpvalone &             \rpvaltwo\\
         ~~~$\rho_{P}$\dotfill &                             Density (cgs)\dotfill &           \rhopvalone &           \rhopvaltwo\\
      ~~~$\log{g_{P}}$\dotfill &                           Surface gravity\dotfill &          \loggpvalone &          \loggpvaltwo\\
           ~~~$T_{eq}$\dotfill &               Equilibrium temperature (K)\dotfill &            \teqvalone &            \teqvaltwo\\
           ~~~$\Theta$\dotfill &                           Safronov number\dotfill &          \thetavalone &          \thetavaltwo\\
            ~~~$\fave$\dotfill &                  Incident flux (\fluxcgs)\dotfill &           \favevalone &           \favevaltwo\\
\sidehead{RV Parameters:}
              ~~~$T_C$\dotfill &    Time of inferior conjunction (\bjdtdb)\dotfill &             \tcvalone &             \tcvaltwo\\
            ~~~$T_{P}$\dotfill &              Time of periastron (\bjdtdb)\dotfill &             \tpvalone &             \tpvaltwo\\
                ~~~$K$\dotfill &                   RV semi-amplitude (m/s)\dotfill &              \kvalone &              \kvaltwo\\
              ~~~$K_R$\dotfill &                        RM amplitude (m/s)\dotfill &             \krvalone &             \krvaltwo\\
       ~~~$M_P\sin{i}$\dotfill &                        Minimum mass (\mj)\dotfill &         \mpsinivalone &         \mpsinivaltwo\\
      ~~~$M_{P}/M_{*}$\dotfill &                                Mass ratio\dotfill &        \mpmstarvalone &        \mpmstarvaltwo\\
                ~~~$u$\dotfill &                  RM linear limb darkening\dotfill &              \uvalone &              \uvaltwo\\
  ~~~$\gamma_{EXPERT}$\dotfill &                                       m/s\dotfill &    \gammaexpertvalone &    \gammaexpertvaltwo\\
    ~~~$\gamma_{FIES}$\dotfill &                                       m/s\dotfill &      \gammafiesvalone &      \gammafiesvaltwo\\
   ~~~$\gamma_{HIRES}$\dotfill &                                       m/s\dotfill &     \gammahiresvalone &     \gammahiresvaltwo\\
    ~~~$\gamma_{TRES}$\dotfill &                                       m/s\dotfill &      \gammatresvalone &      \gammatresvaltwo\\
     ~~~$\dot{\gamma}$\dotfill &                        RV slope (m/s/day)\dotfill &       \dotgammavalone &       \dotgammavaltwo\\  
           ~~~$\ecosw$\dotfill &                                          \dotfill &          \ecoswvalone &          \ecoswvaltwo\\
           ~~~$\esinw$\dotfill &                                          \dotfill &          \esinwvalone &          \esinwvaltwo\\
         ~~~$f(m1,m2)$\dotfill &                       Mass function (\mj)\dotfill &      \fmonemtwovalone &      \fmonemtwovaltwo\\
  ~~~$\sigma_{EXPERT}$\dotfill &                  Error scaling for EXPERT\dotfill &                  2.00 &                 1.37 \\
    ~~~$\sigma_{FIES}$\dotfill &                    Error scaling for FIES\dotfill &                  1.00 &                 1.00 \\
   ~~~$\sigma_{HIRES}$\dotfill &                   Error scaling for HIRES\dotfill &                  7.48 &                 6.85 \\
    ~~~$\sigma_{TRES}$\dotfill &                    Error scaling for TRES\dotfill &                  2.32 &                 2.09 \\
\sidehead{Primary Transit Parameters:}
      ~~~$R_{P}/R_{*}$\dotfill &     Radius of the planet in stellar radii\dotfill &        \rprstarvalone &        \rprstarvaltwo\\
            ~~~$a/R_*$\dotfill &          Semi-major axis in stellar radii\dotfill &             \arvalone &             \arvaltwo\\
                ~~~$i$\dotfill &                     Inclination (degrees)\dotfill &              \ivalone &              \ivaltwo\\
                ~~~$b$\dotfill &                          Impact parameter\dotfill &              \bvalone &              \bvaltwo\\
           ~~~$\delta$\dotfill &                             Transit depth\dotfill &          \deltavalone &          \deltavaltwo\\
         ~~~$T_{FWHM}$\dotfill &                      FWHM duration (days)\dotfill &          \tfwhmvalone &          \tfwhmvaltwo\\
             ~~~$\tau$\dotfill &            Ingress/egress duration (days)\dotfill &            \tauvalone &            \tauvaltwo\\
           ~~~$T_{14}$\dotfill &                     Total duration (days)\dotfill &       \tonefourvalone &       \tonefourvaltwo\\
            ~~~$P_{T}$\dotfill &  A priori non-grazing transit probability\dotfill &             \ptvalone &             \ptvaltwo\\
          ~~~$P_{T,G}$\dotfill &               A priori transit probability\dotfill &            \ptgvalone &            \ptgvaltwo\\
\sidehead{Secondary Eclipse Parameters:}
            ~~~$T_{S}$\dotfill &                 Time of eclipse (\bjdtdb)\dotfill &             \tsvalone &             \tsvaltwo\\
            ~~~$b_{S}$\dotfill &                          Impact parameter\dotfill &             \bsvalone &             \bsvaltwo\\
       ~~~$T_{S,FWHM}$\dotfill &                      FWHM duration (days)\dotfill &         \tsfwhmvalone &         \tsfwhmvaltwo\\
           ~~~$\tau_S$\dotfill &            Ingress/egress duration (days)\dotfill &           \tausvalone &           \tausvaltwo\\
         ~~~$T_{S,14}$\dotfill &                     Total duration (days)\dotfill &      \tsonefourvalone &      \tsonefourvaltwo\\
            ~~~$P_{S}$\dotfill &  A priori non-grazing eclipse probability\dotfill &             \psvalone &             \psvaltwo\\
          ~~~$P_{S,G}$\dotfill &              A priori eclipse probability\dotfill &            \psgvalone &            \psgvaltwo\\
\enddata
\label{tab:exofastpars}
\end{deluxetable*}

\hyphenation{time-stamps}
\subsection{Transit Timing Variations}
\begin{figure}[!htbp]
  \begin{center}
    \includegraphics[width=3.25in]{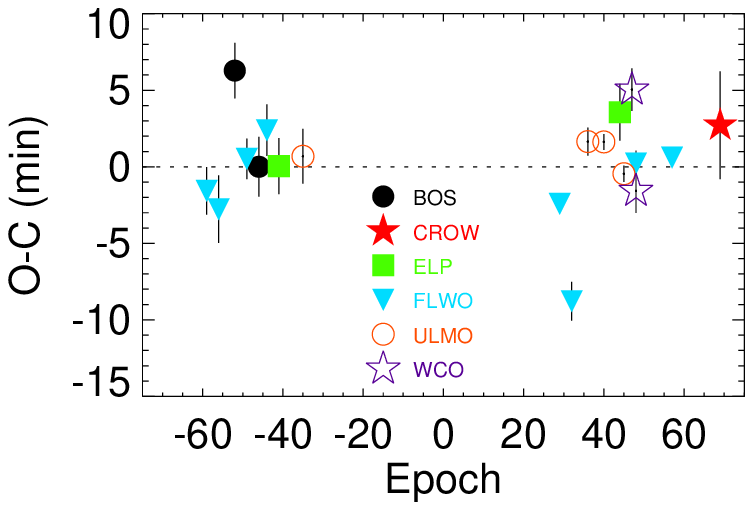} 
    \caption{The transit times of the 19 transits of \pname \ with the
      best-fit linear ephemeris ($T_{0} = $\ttvaltwo, $P=$\pervaltwo) subtracted.}
    \label{fig:ttvs}
  \end{center}
\end{figure}

Great care was taken to translate each of our timestamps to a common
system, \bjdtdb \ \citep{eastman10b}. All observers report \jdutc \ at mid
exposure and the translation to \bjdtdb \ is done uniformly for all
observations prior to the fit. In addition, we have double checked the
values quoted directly from an example image header for each
observer. During the global fit, the transit time for each of the 19
transits were allowed to vary freely, as shown in Figure
\ref{fig:ttvs} and summarized in Table \ref{tab:ttvs}. While most
epochs were consistent with a linear ephemeris,

\begin{equation}
T_{0} = \ttvaltwo, P=\pervaltwo, 
\end{equation}

\noindent there are a few large outliers. However, given the TTV results for Hot Jupiters from the Kepler mission \citep{steffen12}, the
heterogeneity of our clocks, observatories, and observing procedures,
and the potential for atmospheric and astrophysical sources of red
noise to skew our transit times by amounts larger than our naive error
estimates imply \citep{carter09}, we do not view these outliers as
significant. In particular, our experience with KELT-3b
\citep{pepper13}, where we observed the same epoch with 3 different
telescopes and found that the observation from FLWO differed by
5-sigma (7 minutes) from the other two with no discernible cause has
led us to be skeptical of all ground-based TTV detections. Curiously,
the two most significant outliers are also from FLWO, possibly
pointing to a problem with the stability of its observatory clock (at
the 5-10 minute level). We have set up monitoring of this clock and are watching it closely both for drifts and short-term glitches. While we feel our skepticism of these nominally significant TTVs is warranted, the recent results for WASP-47b \citep{becker15} is a counter-example to the observation that Hot Jupiters tend not to have companions, so these outliers may be worth additional follow-up with a more homogeneous setup.

\begin{deluxetable*}{lccccccc}[!htbp]
\tablecaption{Median values and 68\% confidence interval for the transit times of all 19 follow-up light curves of \pname \ from the global circular fit, along with residuals from the best-fit linear ephemeris: $T_{C,N} (\bjdtdb) = \ttvaltwo + N\left(\pervaltwo\right)$}
\tablehead{\colhead{Parameter} & \colhead{UT Date} & \colhead{Telescope} & \colhead{Filter} & \colhead{Epoch} & \colhead{$T_C$ (\bjdtdb)} & \colhead{O-C} (sec) & \colhead{O-C ($\sigma_{T_C}$)}}

\startdata
$T_{C,0}$\dotfill& 2012-03-30 & FLWO  & $i$   & -59 & \tczerovaltwo     &   -94.08 & -1.01\\
$T_{C,1}$\dotfill& 2012-04-08 & FLWO  & $g$   & -56 & \tconevaltwo      &  -166.03 & -1.25\\
$T_{C,2}$\dotfill& 2012-04-20 & BOS   & $i$   & -52 & \tctwovaltwo      &   377.05 &  3.46\\
$T_{C,3}$\dotfill& 2012-04-29 & FLWO  & $z$   & -49 & \tcthreevaltwo    &    31.30 &  0.39\\
$T_{C,4}$\dotfill& 2012-05-08 & BOS   & $g$   & -46 & \tcfourvaltwo     &     0.14 &  0.00\\
$T_{C,5}$\dotfill& 2012-05-14 & FLWO  & $g$   & -44 & \tcfivevaltwo     &   143.37 &  1.42\\
$T_{C,6}$\dotfill& 2012-05-23 & ELP   & $g$   & -41 & \tcsixvaltwo      &     2.65 &  0.02\\
$T_{C,7}$\dotfill& 2012-06-10 & ULMO  & $r$   & -35 & \tcsevenvaltwo    &    41.23 &  0.39\\ 
$T_{C,8}$\dotfill& 2012-12-18 & FLWO  & $i$   &  29 & \tceightvaltwo    &  -145.14 & -4.44\\
$T_{C,9}$\dotfill& 2012-12-27 & FLWO  & $i$   &  32 & \tcninevaltwo     &  -527.01 & -6.92\\
$T_{C,10}$\dotfill& 2013-01-08 & ULMO & $g$   &  36 & \tconezerovaltwo  &    98.93 &  1.80\\
$T_{C,11}$\dotfill& 2013-01-20 & ULMO & $g$   &  40 & \tconeonevaltwo   &    97.52 &  3.14\\
$T_{C,12}$\dotfill& 2013-02-01 & ELP  & $z$   &  44 & \tconetwovaltwo   &   213.79 &  1.93\\
$T_{C,13}$\dotfill& 2013-02-04 & ULMO & $g$   &  45 & \tconethreevaltwo &   -27.14 & -0.85\\
$T_{C,14}$\dotfill& 2013-02-10 & WCO  & $CBB$ &  47 & \tconefourvaltwo  &   302.63 &  3.61\\
$T_{C,15}$\dotfill& 2013-02-13 & WCO  & $CBB$ &  48 & \tconefivevaltwo  &   -94.51 & -1.10\\
$T_{C,16}$\dotfill& 2013-02-13 & FLWO & $z$   &  48 & \tconesixvaltwo   &    13.23 &  0.26\\
$T_{C,17}$\dotfill& 2013-03-12 & FLWO & $i$   &  57 & \tconesevenvaltwo &    35.76 &  1.01\\
$T_{C,18}$\dotfill& 2013-04-17 & CROW & $I$   &  69 & \tconeeightvaltwo &   162.86 &  0.77\\
\enddata
\label{tab:ttvs}
\end{deluxetable*}

\section{False Positive Rejection}

False positives due to background eclipsing binaries are common in
transit surveys. As such, all KELT candidates are subject to a
rigorous set of tests to eliminate such scenarios. While our AO
results show that our survey data and followup photometry were diluted
by a companion binary system, KELT-4A was resolved in all of the
radial velocity observations used for analysis, which shows a clear signal of
a planet. In addition, there was no evidence of any other background
stars in the area (see \S \ref{sec:ao}). We also observed transits of
\pname \ in six different filters to check for a wavelength-dependent
transit depth indicative of a blend, but all transit depths in all
bands were consistent with one another after accounting for the blend
with the nearby companion.

\section{Insolation Evolution}

\begin{figure}
  \begin{center}
    \includegraphics[width=2.5in,angle=90]{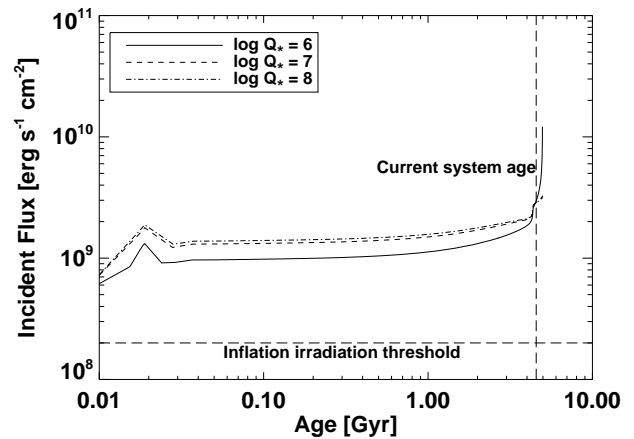} 
    \caption{Change in incident flux for \pname, with different test
      values for $Q_*$ for \starname. In all cases the planet has
      always received more than enough flux from its host to keep the
      planet irradiated beyond the insolation threshold of $2 \times
      10^8 \ erg \ s^{-1} cm^{-2}$ identified by \citet{demory11}.}
    \label{fig:irad}
  \end{center}
\end{figure}

Because \pname \ is inflated, it is interesting to investigate its irradiation history, as described in \citet{pepper13}, as an empirical probe into the timescale of inflation mechanisms \citep{assef09,spiegel12}. Our results are shown in Figure \ref{fig:irad}. Similar to KELT-3b, the incident flux has always been above the inflation irradiation threshold identified by \citet{demory11}, regardless of our assumptions about the tidal Q factor. Similar to KELT-8b, it is likely spiraling into its host star with all reasonable values of the tidal Q factor \citep{fulton15}.

For this model, we matched the current conditions at the age of KELT-4A. However, we note that instead of using the YY stellar models as in the rest of the analysis, we used the YREC models \citep{siess00, demarque08} here. As a result, we could not precisely match the stellar parameters used elsewhere in the modeling, but they were well within the quoted uncertainties.

\section{Discussion}

The large separation of the planet host from the tight binary makes this system qualitatively similar to KELT-2Ab \citep{beatty12}. As such, we expect the Kozai mechanism \citep{kozai62, lidov62} to influence the migration of \pname \ as well, and therefore the \pname \ system to be misaligned. With an expected RM amplitude of $50~{\rm m~s^{-1}}$, this would be easy to detect, though complicated by its near-integer day period and nearby companion. It is well-positioned for RM observations at Keck in 2016. Like KELT-2A, the effective temperature of KELT-4A (\teffvaltwo \ K) is near the dividing line between cool aligned stars and hot misaligned stars noted by \citet{winn10c}.

Interestingly, relative periods discussed in \S \ref{sec:ao} set a Kozai-Lidov (KL) timescale of 540,000 years for the KELT-4BC stellar binary \citep{pejcha13}. This is relatively short, so we may expect to find that BC is currently undergoing Kozai-Lidov cycles and therefore be highly eccentric. Its relatively short period makes this an excellent candidate for continued follow-up effort, though its current separation is right at the K-band diffraction limit for Keck (25.5 mas).

The planetary binary has a Kozai-Lidov timescale of 1.6 Gyr. This is well below the age of the system, but assuming \pname \ formed beyond the snow line ($\sim5$ AU), its period would have been longer and therefore its Kozai-Lidov timescale shorter. This suggests Kozai is a plausible migration mechanism.

\begin{figure*}
\centering
\includegraphics[width=0.95\textwidth]{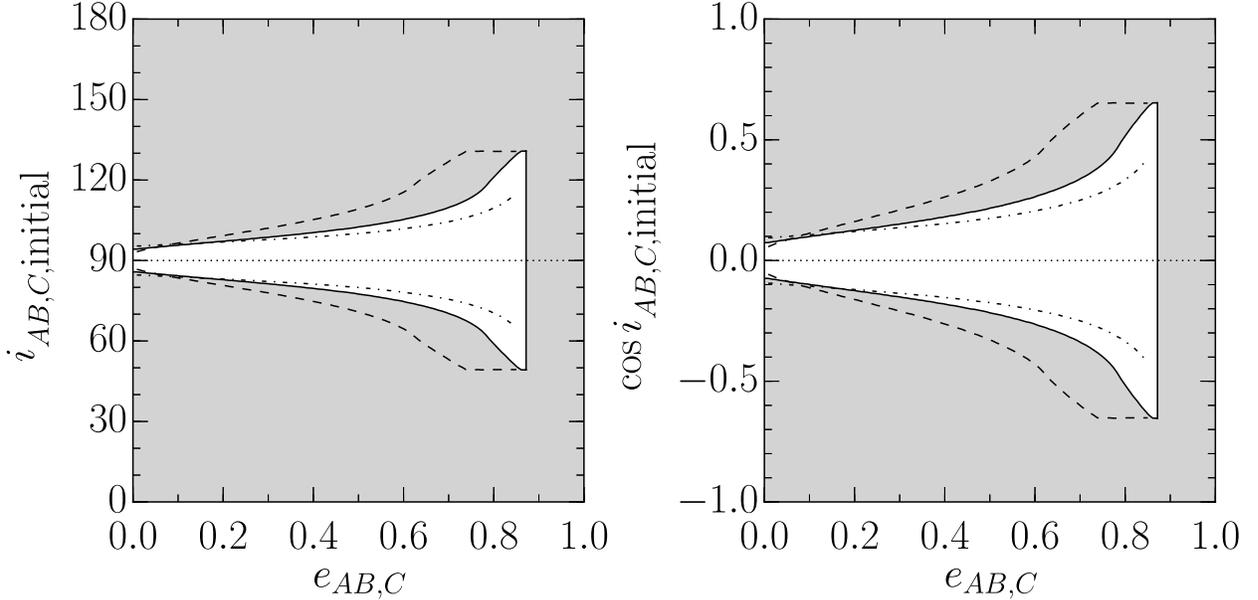}

\caption{Ranges of initial inclination and outer eccentricity for which Kozai-Lidov oscillations can drive \pname \ to its present semi-major axis, starting from a semi-major axis of 3 AU (dot-dashed line), 5 AU (solid line), and 10 AU (dashed line).  The upper limit on the eccentricity is due to the requirement that the system be dynamically stable.  We show in the right panel the same constraints, but with the cosine of the initial inclination rather than the inclination itself.  Because $\cos i$ is expected to be uniformly distributed in hierarchical triples and the distribution of eccentricities of wide binaries is also observed to be approximately uniform, equal areas of this plot may be interpreted as equal probabilities.}

\label{fig:pspace}
\end{figure*}

If \pname \ formed past the ice line at a few AU and migrated to its present location via Kozai-Lidov oscillations and tidal friction \citep[as in, e.g.,][]{wu03, fabrycky07}, this would place constraints on the orbital parameters of the system as it existed shortly after formation.  In particular, for Kozai-Lidov oscillations to be strong enough to drive the planet from $\sim$5 AU to 0.04 AU either the initial inclination of the outer orbit relative to the planet must have been close to 90$^{\circ}$, or the eccentricity of the outer orbit must have been large, or both.  We quantify these constraints by using the \texttt{kozai} Python package \citep{antognini15} to evolve a set of hierarchical triples in the secular approximation with the observed orbital parameters, but varying the outer eccentricity and the mutual inclination between the planetary orbit and the outer orbit.  Although the KELT-4 system contains four bodies, we take the KELT-4BC system to be a point mass of 1.3 $M_{\odot}$.  Combinations of inclination and outer eccentricity that can drive strong enough Kozai-Lidov oscillations to bring the planet to within 0.02 AU\footnote{the planet circularizes at twice the initial periastron distance due to conservation of angular momentum \citep{fabrycky07}} of the star are shown in the unshaded region of Figure \ref{fig:pspace}.  Since the distribution of $\cos i$ between the planetary orbit and the outer orbit is expected to be uniform and the eccentricity distribution of wide binaries is also observed to be approximately uniform, equal areas of the right panel of Figure \ref{fig:pspace} can be interpreted as equal probabilities.  Although we did not include relativistic precession in these calculations, we compared the precession timescale to the period of the KL oscillations and found that the precession timescale was much longer (at least a factor of 10) in all cases in which the KL oscillations are strong enough to drive \pname \ to its present location.

While there are several planets in binary stellar systems, there are only a few transiting planets known in hierarchical triples.  These systems may have had a richer dynamical history than the more commonly found planets in binary systems.  \citet{pejcha13} and \citet{hamers15} have found that in quadruple systems the presence of the additional body can, in some cases, lead to resonant interactions between the Kozai-Lidov oscillations that occur in the inner binaries, thereby producing stronger eccentricity oscillations than in hierarchical triples with similar orbital parameters. Due to the small mass of \pname \ it would not have had any strong dynamical influence on KELT-4BC, but the binarity of KELT-4BC may have influenced the dynamical evolution of \pname.  The discovery of systems like KELT-4 highlights the need for further study of the dynamics of quadruple systems.

High-resolution imaging and RV monitoring of both KELT-4A and KELT-4BC is likely to constrain the orbit of the twins relatively well in a short time. Their inclination and eccentricity is likely to provide insight into the formation and dynamical evolution of the system. The very long period of KELT-4BC around KELT-4A makes characterizing the orbit of the KELT-4BC binary around the KELT-4A primary more challenging, but continued monitoring may be able to exclude certain inclinations or eccentricities.

We expect, in the three years since the original AO observations, motions of $\sim37$ mas between KELT-4B and KELT-4C, which should be easily visible. Between KELT-4A and KELT-4BC, the motions are expected to be $\sim10$ mas, which should also be marginally detectable with additional Keck observations. {\it Gaia} \citep{perryman01} will provide new absolute astrometric measurements on KELT-4A to $\sim7 \mu as$ and an unresolved position of KELT-4BC to $\sim25 \mu as$. With $\sim25 \mu as$ accuracy, {\it Gaia} could see significant motion of KELT-4BC in $\sim3$ days. {\it Gaia} will also provide a precise distance which will allow us to infer the radius of the primary, thereby distinguishing between the marginally inconsistent stellar parameters.

The maximum RV semi-amplitude of KELT-4A induced by the KELT-BC system, (i.e., assuming an edge-on orbit), would be $1.3~{\rm km~s^{-1}}$ or $0.002~{\rm m~s^{-1}~day^{-1}}$. Since a separate zero point is fit for each data set, we are not sensitive to secular drifts that span the entire data set. We allowed the slope to be free during the fit, but the uncertainty is a factor of 20 larger than the expected signal. Still, it is not unrealistic to expect to detect a drift with long-term monitoring. The KELT-4BC system would have a maximum semi-amplitude of $5.3~{\rm km~s^{-1}}$ or about $2~{\rm m~s^{-1}~day^{-1}}$, which is easily detectable, though its $V=13$ magnitude and 1.5" separation from the primary star make it a challenging target.

\begin{figure}[!htbp]
  \begin{center}
    \includegraphics[width=3.25in]{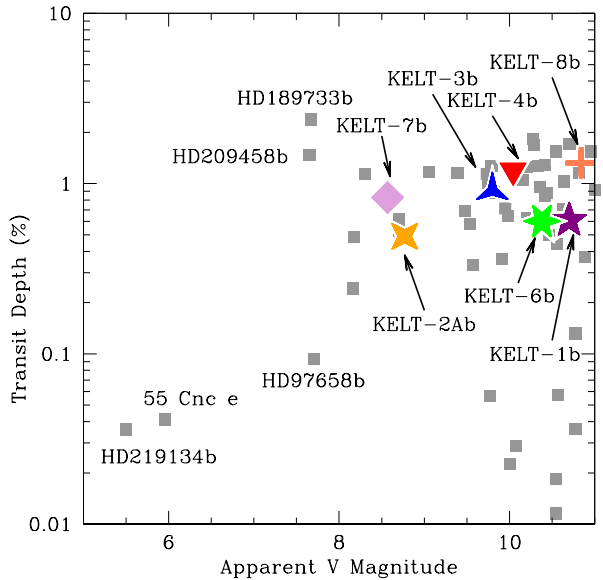} 
    \caption{Transit depth as a function of the apparent V magnitude
      of the host star for a sample of transiting systems. \pname \ is
      shown as the red triangle. All else being equal, objects in the
      top left provide the best targets for follow-up. The other
      discoveries from the KELT survey are also shown.}
    \label{fig:rprsv}
  \end{center}
\end{figure}

This inflated hot Jupiter, while not unique (e.g., HAT-P-39b, HAT-P-40b, HAT-P-41b \citep{hartman12}), like all KELT planets, is among the brightest and therefore easiest to follow up as a result of our survey design (see Figure \ref{fig:rprsv}). In particular, high-resolution imaging capable of resolving the stellar binary (42 mas) would help constrain the orbit of KELT-4BC, and may help create a more robust migration history of the entire system.

\acknowledgements

We thank Geoff Marcy for use of his Keck/HIRES time and doing observations. We extend special thanks to those of Hawai'ian ancestry on whose sacred mountain of Mauna Kea we are privileged to be guests.  Without their generous hospitality, the Keck observations presented herein would not have been possible. Early work on KELT-North was supported by NASA Grant NNG04GO70G. Work by B.S.G., J.D.E., and T.G.B. was partially supported by NSF CAREER Grant AST-1056524. K.A.C. was supported by a NASA Kentucky Space Grant Consortium Graduate Fellowship. J.A.P. and K.G.S. acknowledge support from the Vanderbilt Office of the Provost through the Vanderbilt Initiative in Data-intensive Astrophysics. K.G.S. and L.H. acknowledge the support of the National Science Foundation through PAARE grant AST-0849736 and AAG grant AST-1009810. The TRES and KeplerCam observations were obtained with partial support from the Kepler Mission through NASA Cooperative Agreement NNX11AB99A with the Smithsonian Astrophysical Observatory, D.W.L. PI. J. M. O. A. is supported in part by NSF Award \#1313252. J.G., B.M., and J.W. acknowledge support from NSF AST-0705139 and the University of Florida for the development of the EXPERT instrument and observations. This material is based upon work supported by the National Science Foundation Graduate Research Fellowship under grant No. 2014184874. Any opinion, findings, and conclusions or recommendations expressed in this material are those of the authors(s) and do not necessarily reflect the views of the National Science Foundation.

The Byrne Observatory at Sedgwick (BOS) is operated by the Las Cumbres Observatory Global Telescope Network and is located at the Sedgwick Reserve, a part of the University of California Natural Reserve System. This work makes use of observations from the LCOGT network.

This work has made use of NASA's Astrophysics Data System, the Extrasolar Planet Encyclopedia at exoplanet.eu \citep{schneider11}, the SIMBAD database operated at CDS, Strasbourg, France, and the VizieR catalogue access tool, CDS, Strasbourg, France \citep{ochsenbein00}. Certain calculations in this paper were carried out on the Ruby cluster operated by the Ohio Supercomputer Center \citep{ruby15}.

This publication makes use of data products from the Wide-field Infrared Survey Explorer, which is a joint project of the University of California, Los Angeles, and the Jet Propulsion Laboratory/California Institute of Technology, funded by the National Aeronautics and Space Administration.

This publication makes use of data products from the Two Micron All Sky Survey, which is a joint project of the University of Massachusetts and the Infrared Processing and Analysis Center/California Institute of Technology, funded by the National Aeronautics and Space Administration and the National Science Foundation.

This paper makes use of data from the first public release of the WASP data \citep{butters10} as provided by the WASP consortium and services at the NASA Exoplanet Archive \citep{akeson13}, which is operated by the California Institute of Technology, under contract with the National Aeronautics and Space Administration under the Exoplanet Exploration Program, the Exoplanet Orbit Database and the Exoplanet Data Explorer at exoplanets.org \citep{wright11}.


\begin{thebibliography}{100}
\expandafter\ifx\csname natexlab\endcsname\relax\def\natexlab#1{#1}\fi

\bibitem[{{Adams} \& {Laughlin}(2006)}]{adams06}
{Adams}, F.~C., \& {Laughlin}, G. 2006, \apj, 649, 1004

\bibitem[{{Akeson} {et~al.}(2013){Akeson}, {Chen}, {Ciardi}, {Crane}, {Good},
  {Harbut}, {Jackson}, {Kane}, {Laity}, {Leifer}, {Lynn}, {McElroy}, {Papin},
  {Plavchan}, {Ram{\'{\i}}rez}, {Rey}, {von Braun}, {Wittman}, {Abajian},
  {Ali}, {Beichman}, {Beekley}, {Berriman}, {Berukoff}, {Bryden}, {Chan},
  {Groom}, {Lau}, {Payne}, {Regelson}, {Saucedo}, {Schmitz}, {Stauffer},
  {Wyatt}, \& {Zhang}}]{akeson13}
{Akeson}, R.~L., {et~al.} 2013, \pasp, 125, 989

\bibitem[{{Alonso} {et~al.}(2004){Alonso}, {Brown}, {Torres}, {Latham},
  {Sozzetti}, {Mandushev}, {Belmonte}, {Charbonneau}, {Deeg}, {Dunham},
  {O'Donovan}, \& {Stefanik}}]{alonso04}
{Alonso}, R., {et~al.} 2004, \apjl, 613, L153

\bibitem[{{Antognini}(2015)}]{antognini15}
{Antognini}, J.~M.~O. 2015, \mnras, 452, 3610

\bibitem[{{Argue} {et~al.}(1992){Argue}, {Bunclark}, {Irwin}, {Lampens},
  {Sinachopoulos}, \& {Wayman}}]{argue92}
{Argue}, A.~N., {Bunclark}, P.~S., {Irwin}, M.~J., {Lampens}, P.,
  {Sinachopoulos}, D., \& {Wayman}, P.~A. 1992, \mnras, 259, 563

\bibitem[{{Assef} {et~al.}(2009){Assef}, {Gaudi}, \& {Stanek}}]{assef09}
{Assef}, R.~J., {Gaudi}, B.~S., \& {Stanek}, K.~Z. 2009, \apj, 701, 1616

\bibitem[{{Baglin} {et~al.}(2006){Baglin}, {Auvergne}, {Boisnard}, {Lam-Trong},
  {Barge}, {Catala}, {Deleuil}, {Michel}, \& {Weiss}}]{baglin06}
{Baglin}, A., {et~al.} 2006, in COSPAR, Plenary Meeting, Vol.~36, 36th COSPAR
  Scientific Assembly, 3749--+

\bibitem[{{Bakos} {et~al.}(2002){Bakos}, {L{\'a}z{\'a}r}, {Papp}, {S{\'a}ri},
  \& {Green}}]{bakos02}
{Bakos}, G.~{\'A}., {L{\'a}z{\'a}r}, J., {Papp}, I., {S{\'a}ri}, P., \&
  {Green}, E.~M. 2002, \pasp, 114, 974

\bibitem[{{Beatty} \& {Gaudi}(2008)}]{beatty08}
{Beatty}, T.~G., \& {Gaudi}, B.~S. 2008, \apj, 686, 1302

\bibitem[{{Beatty} {et~al.}(2012){Beatty}, {Pepper}, {Siverd}, {Eastman},
  {Bieryla}, {Latham}, {Buchhave}, {Jensen}, {Manner}, {Stassun}, {Gaudi},
  {Berlind}, {Calkins}, {Collins}, {DePoy}, {Esquerdo}, {Fulton}, {F{\H
  u}r{\'e}sz}, {Geary}, {Gould}, {Hebb}, {Kielkopf}, {Marshall}, {Pogge},
  {Stanek}, {Stefanik}, {Street}, {Szentgyorgyi}, {Trueblood}, {Trueblood}, \&
  {Stutz}}]{beatty12}
{Beatty}, T.~G., {et~al.} 2012, \apjl, 756, L39

\bibitem[{{Bechter} {et~al.}(2014){Bechter}, {Crepp}, {Ngo}, {Knutson},
  {Batygin}, {Hinkley}, {Muirhead}, {Johnson}, {Howard}, {Montet}, {Matthews},
  \& {Morton}}]{bechter14}
{Bechter}, E.~B., {et~al.} 2014, \apj, 788, 2

\bibitem[{{Becker} {et~al.}(2015){Becker}, {Vanderburg}, {Adams}, {Rappaport},
  \& {Schwengeler}}]{becker15}
{Becker}, J.~C., {Vanderburg}, A., {Adams}, F.~C., {Rappaport}, S.~A., \&
  {Schwengeler}, H.~M. 2015, ArXiv e-prints

\bibitem[{{Bensby} {et~al.}(2003){Bensby}, {Feltzing}, \&
  {Lundstr{\"o}m}}]{bensby03}
{Bensby}, T., {Feltzing}, S., \& {Lundstr{\"o}m}, I. 2003, \aap, 410, 527

\bibitem[{{Brown} {et~al.}(2013){Brown}, {Baliber}, {Bianco}, {Bowman},
  {Burleson}, {Conway}, {Crellin}, {Depagne}, {De Vera}, {Dilday}, {Dragomir},
  {Dubberley}, {Eastman}, {Elphick}, {Falarski}, {Foale}, {Ford}, {Fulton},
  {Garza}, {Gomez}, {Graham}, {Greene}, {Haldeman}, {Hawkins}, {Haworth},
  {Haynes}, {Hidas}, {Hjelstrom}, {Howell}, {Hygelund}, {Lister}, {Lobdill},
  {Martinez}, {Mullins}, {Norbury}, {Parrent}, {Paulson}, {Petry}, {Pickles},
  {Posner}, {Rosing}, {Ross}, {Sand}, {Saunders}, {Shobbrook}, {Shporer},
  {Street}, {Thomas}, {Tsapras}, {Tufts}, {Valenti}, {Vander Horst}, {Walker},
  {White}, \& {Willis}}]{brown13}
{Brown}, T.~M., {et~al.} 2013, \pasp, 125, 1031

\bibitem[{{Buchhave} {et~al.}(2010){Buchhave}, {Bakos}, {Hartman}, {Torres},
  {Kov{\'a}cs}, {Latham}, {Noyes}, {Esquerdo}, {Everett}, {Howard}, {Marcy},
  {Fischer}, {Johnson}, {Andersen}, {F{\H u}r{\'e}sz}, {Perumpilly},
  {Sasselov}, {Stefanik}, {B{\'e}ky}, {L{\'a}z{\'a}r}, {Papp}, \&
  {S{\'a}ri}}]{buchhave10}
{Buchhave}, L.~A., {et~al.} 2010, \apj, 720, 1118

\bibitem[{{Buchhave} {et~al.}(2012){Buchhave}, {Latham}, {Johansen},
  {Bizzarro}, {Torres}, {Rowe}, {Batalha}, {Borucki}, {Brugamyer}, {Caldwell},
  {Bryson}, {Ciardi}, {Cochran}, {Endl}, {Esquerdo}, {Ford}, {Geary},
  {Gilliland}, {Hansen}, {Isaacson}, {Laird}, {Lucas}, {Marcy}, {Morse},
  {Robertson}, {Shporer}, {Stefanik}, {Still}, \& {Quinn}}]{buchhave12}
---. 2012, \nat, 486, 375

\bibitem[{{Buchhave} {et~al.}(2014){Buchhave}, {Bizzarro}, {Latham},
  {Sasselov}, {Cochran}, {Endl}, {Isaacson}, {Juncher}, \&
  {Marcy}}]{buchhave14}
---. 2014, \nat, 509, 593

\bibitem[{{Butters} {et~al.}(2010){Butters}, {West}, {Anderson}, {Collier
  Cameron}, {Clarkson}, {Enoch}, {Haswell}, {Hellier}, {Horne}, {Joshi},
  {Kane}, {Lister}, {Maxted}, {Parley}, {Pollacco}, {Smalley}, {Street},
  {Todd}, {Wheatley}, \& {Wilson}}]{butters10}
{Butters}, O.~W., {et~al.} 2010, \aap, 520, L10

\bibitem[{{Carter} \& {Winn}(2009)}]{carter09}
{Carter}, J.~A., \& {Winn}, J.~N. 2009, \apj, 704, 51

\bibitem[{Center(2015)}]{ruby15}
Center, O.~S. 2015

\bibitem[{{Charbonneau} {et~al.}(2000){Charbonneau}, {Brown}, {Latham}, \&
  {Mayor}}]{charbonneau00}
{Charbonneau}, D., {Brown}, T.~M., {Latham}, D.~W., \& {Mayor}, M. 2000, \apjl,
  529, L45

\bibitem[{{Charbonneau} {et~al.}(2002){Charbonneau}, {Brown}, {Noyes}, \&
  {Gilliland}}]{charbonneau02}
{Charbonneau}, D., {Brown}, T.~M., {Noyes}, R.~W., \& {Gilliland}, R.~L. 2002,
  \apj, 568, 377

\bibitem[{{Charbonneau} {et~al.}(2006){Charbonneau}, {Winn}, {Latham}, {Bakos},
  {Falco}, {Holman}, {Noyes}, {Cs{\'a}k}, {Esquerdo}, {Everett}, \&
  {O'Donovan}}]{charbonneau06}
{Charbonneau}, D., {et~al.} 2006, \apj, 636, 445

\bibitem[{{Chubak} {et~al.}(2012){Chubak}, {Marcy}, {Fischer}, {Howard},
  {Isaacson}, {Johnson}, \& {Wright}}]{chubak12}
{Chubak}, C., {Marcy}, G., {Fischer}, D.~A., {Howard}, A.~W., {Isaacson}, H.,
  {Johnson}, J.~A., \& {Wright}, J.~T. 2012, ArXiv e-prints

\bibitem[{{Claret} \& {Bloemen}(2011)}]{claret11}
{Claret}, A., \& {Bloemen}, S. 2011, \aap, 529, A75+

\bibitem[{{Co{\c s}kuno{\v g}lu} {et~al.}(2011){Co{\c s}kuno{\v g}lu}, {Ak},
  {Bilir}, {Karaali}, {Yaz}, {Gilmore}, {Seabroke}, {Bienaym{\'e}},
  {Bland-Hawthorn}, {Campbell}, {Freeman}, {Gibson}, {Grebel}, {Munari},
  {Navarro}, {Parker}, {Siebert}, {Siviero}, {Steinmetz}, {Watson}, {Wyse}, \&
  {Zwitter}}]{coskunoglu11}
{Co{\c s}kuno{\v g}lu}, B., {et~al.} 2011, \mnras, 412, 1237

\bibitem[{{Collier Cameron} {et~al.}(2007){Collier Cameron}, {Bouchy},
  {H{\'e}brard}, {Maxted}, {Pollacco}, {Pont}, {Skillen}, {Smalley}, {Street},
  {West}, {Wilson}, {Aigrain}, {Christian}, {Clarkson}, {Enoch}, {Evans},
  {Fitzsimmons}, {Fleenor}, {Gillon}, {Haswell}, {Hebb}, {Hellier}, {Hodgkin},
  {Horne}, {Irwin}, {Kane}, {Keenan}, {Loeillet}, {Lister}, {Mayor}, {Moutou},
  {Norton}, {Osborne}, {Parley}, {Queloz}, {Ryans}, {Triaud}, {Udry}, \&
  {Wheatley}}]{cameron07}
{Collier Cameron}, A., {et~al.} 2007, \mnras, 375, 951

\bibitem[{{Collins} \& {Kielkopf}(2013)}]{collins13b}
{Collins}, K., \& {Kielkopf}, J. 2013, {AstroImageJ: ImageJ for Astronomy},
  Astrophysics Source Code Library

\bibitem[{{Collins}(2015)}]{collins15}
{Collins}, K.~A. 2015, Electronic Theses and Dissertations, Paper 2104

\bibitem[{{Collins} {et~al.}(2014){Collins}, {Eastman}, {Beatty}, {Siverd},
  {Gaudi}, {Pepper}, {Kielkopf}, {Johnson}, {Howard}, {Fischer}, {Manner},
  {Bieryla}, {Latham}, {Fulton}, {Gregorio}, {Buchhave}, {Jensen}, {Stassun},
  {Penev}, {Crepp}, {Hinkley}, {Street}, {Cargile}, {Mack}, {Oberst}, {Avril},
  {Mellon}, {McLeod}, {Penny}, {Stefanik}, {Berlind}, {Calkins}, {Mao},
  {Richert}, {DePoy}, {Esquerdo}, {Gould}, {Marshall}, {Oelkers}, {Pogge},
  {Trueblood}, \& {Trueblood}}]{collins14}
{Collins}, K.~A., {et~al.} 2014, \aj, 147, 39

\bibitem[{{Couteau}(1973)}]{couteau73}
{Couteau}, P. 1973, \aaps, 10, 273

\bibitem[{{Cutri} \& {et al.}(2012)}]{cutri12}
{Cutri}, R.~M., \& {et al.} 2012, VizieR Online Data Catalog, 2311, 0

\bibitem[{{Cutri} {et~al.}(2003){Cutri}, {Skrutskie}, {van Dyk}, {Beichman},
  {Carpenter}, {Chester}, {Cambresy}, {Evans}, {Fowler}, {Gizis}, {Howard},
  {Huchra}, {Jarrett}, {Kopan}, {Kirkpatrick}, {Light}, {Marsh}, {McCallon},
  {Schneider}, {Stiening}, {Sykes}, {Weinberg}, {Wheaton}, {Wheelock}, \&
  {Zacarias}}]{cutri03}
{Cutri}, R.~M., {et~al.} 2003, VizieR Online Data Catalog, 2246, 0

\bibitem[{{Demarque} {et~al.}(2008){Demarque}, {Guenther}, {Li}, {Mazumdar}, \&
  {Straka}}]{demarque08}
{Demarque}, P., {Guenther}, D.~B., {Li}, L.~H., {Mazumdar}, A., \& {Straka},
  C.~W. 2008, \apss, 316, 31

\bibitem[{{Demarque} {et~al.}(2004){Demarque}, {Woo}, {Kim}, \&
  {Yi}}]{demarque04}
{Demarque}, P., {Woo}, J.-H., {Kim}, Y.-C., \& {Yi}, S.~K. 2004, \apjs, 155,
  667

\bibitem[{{Demory} \& {Seager}(2011)}]{demory11}
{Demory}, B.-O., \& {Seager}, S. 2011, \apjs, 197, 12

\bibitem[{{Demory} {et~al.}(2009){Demory}, {S{\'e}gransan}, {Forveille},
  {Queloz}, {Beuzit}, {Delfosse}, {di Folco}, {Kervella}, {Le Bouquin},
  {Perrier}, {Benisty}, {Duvert}, {Hofmann}, {Lopez}, \& {Petrov}}]{demory09}
{Demory}, B.-O., {et~al.} 2009, \aap, 505, 205

\bibitem[{{Djupvik} \& {Andersen}(2010)}]{djupvik10}
{Djupvik}, A.~A., \& {Andersen}, J. 2010, in Highlights of Spanish Astrophysics
  V, ed. J.~M. {Diego}, L.~J. {Goicoechea}, J.~I. {Gonz{\'a}lez-Serrano}, \&
  J.~{Gorgas}, 211

\bibitem[{{Dragomir} {et~al.}(2013){Dragomir}, {Matthews}, {Eastman},
  {Cameron}, {Howard}, {Guenther}, {Kuschnig}, {Moffat}, {Rowe}, {Rucinski},
  {Sasselov}, \& {Weiss}}]{dragomir13}
{Dragomir}, D., {et~al.} 2013, \apjl, 772, L2

\bibitem[{{Eastman} {et~al.}(2013){Eastman}, {Gaudi}, \& {Agol}}]{eastman13}
{Eastman}, J., {Gaudi}, B.~S., \& {Agol}, E. 2013, \pasp, 125, 83

\bibitem[{{Eastman} {et~al.}(2010){Eastman}, {Siverd}, \& {Gaudi}}]{eastman10b}
{Eastman}, J., {Siverd}, R., \& {Gaudi}, B.~S. 2010, \pasp, 122, 935

\bibitem[{{Fabricius} {et~al.}(2002){Fabricius}, {H{\o}g}, {Makarov}, {Mason},
  {Wycoff}, \& {Urban}}]{fabricius02}
{Fabricius}, C., {H{\o}g}, E., {Makarov}, V.~V., {Mason}, B.~D., {Wycoff},
  G.~L., \& {Urban}, S.~E. 2002, \aap, 384, 180

\bibitem[{{Fabrycky} \& {Tremaine}(2007)}]{fabrycky07}
{Fabrycky}, D., \& {Tremaine}, S. 2007, \apj, 669, 1298

\bibitem[{{F{\H u}r{\'e}sz}(2008)}]{furesz08}
{F{\H u}r{\'e}sz}, G. 2008, PhD Thesis, Univ. Szeged, Hungary

\bibitem[{{Flower}(1996)}]{flower96}
{Flower}, P.~J. 1996, \apj, 469, 355

\bibitem[{{Fortney} {et~al.}(2006){Fortney}, {Saumon}, {Marley}, {Lodders}, \&
  {Freedman}}]{fortney06}
{Fortney}, J.~J., {Saumon}, D., {Marley}, M.~S., {Lodders}, K., \& {Freedman},
  R.~S. 2006, \apj, 642, 495

\bibitem[{{Fulton} {et~al.}(2015){Fulton}, {Collins}, {Gaudi}, {Stassun},
  {Pepper}, {Beatty}, {Siverd}, {Penev}, {Howard}, {Baranec}, {Corfini},
  {Eastman}, {Gregorio}, {Law}, {Lund}, {Oberst}, {Penny}, {Riddle},
  {Rodriguez}, {Stevens}, {Zambelli}, {Ziegler}, {Bieryla}, {D'Ago}, {DePoy},
  {Jensen}, {Kielkopf}, {Latham}, {Manner}, {Marshall}, {McLeod}, \&
  {Reed}}]{fulton15}
{Fulton}, B.~J., {et~al.} 2015, \apj, 810, 30

\bibitem[{{Gaudi} \& {Winn}(2007)}]{gaudi07}
{Gaudi}, B.~S., \& {Winn}, J.~N. 2007, \apj, 655, 550

\bibitem[{{Ge} {et~al.}(2010){Ge}, {Zhao}, {Groot}, {Chang}, {Varosi}, {Wan},
  {Powell}, {Jiang}, {Hanna}, {Wang}, {Pais}, {Liu}, {Dou}, {Schofield},
  {McDowell}, {Costello}, {Delgado-Navarro}, {Fleming}, {Lee}, {Bollampally},
  {Bosman}, {Jakeman}, {Fletcher}, \& {Marquez}}]{ge10}
{Ge}, J., {et~al.} 2010, in Society of Photo-Optical Instrumentation Engineers
  (SPIE) Conference Series, Vol. 7735, Society of Photo-Optical Instrumentation
  Engineers (SPIE) Conference Series

\bibitem[{{Guillot}(2005)}]{guillot05}
{Guillot}, T. 2005, Annual Review of Earth and Planetary Sciences, 33, 493

\bibitem[{{Hamers} {et~al.}(2015){Hamers}, {Perets}, {Antonini}, \& {Portegies
  Zwart}}]{hamers15}
{Hamers}, A.~S., {Perets}, H.~B., {Antonini}, F., \& {Portegies Zwart}, S.~F.
  2015, \mnras, 449, 4221

\bibitem[{{Hartman} {et~al.}(2012){Hartman}, {Bakos}, {B{\'e}ky}, {Torres},
  {Latham}, {Csubry}, {Penev}, {Shporer}, {Fulton}, {Buchhave}, {Johnson},
  {Howard}, {Marcy}, {Fischer}, {Kov{\'a}cs}, {Noyes}, {Esquerdo}, {Everett},
  {Szklen{\'a}r}, {Quinn}, {Bieryla}, {Knox}, {Hinz}, {Sasselov}, {F{\H
  u}r{\'e}sz}, {Stefanik}, {L{\'a}z{\'a}r}, {Papp}, \& {S{\'a}ri}}]{hartman12}
{Hartman}, J.~D., {et~al.} 2012, \aj, 144, 139

\bibitem[{{Hauschildt} {et~al.}(1999){Hauschildt}, {Allard}, \&
  {Baron}}]{hauschildt99}
{Hauschildt}, P.~H., {Allard}, F., \& {Baron}, E. 1999, \apj, 512, 377

\bibitem[{{Henden} {et~al.}(2012){Henden}, {Levine}, {Terrell}, {Smith}, \&
  {Welch}}]{henden12}
{Henden}, A.~A., {Levine}, S.~E., {Terrell}, D., {Smith}, T.~C., \& {Welch}, D.
  2012, Journal of the American Association of Variable Star Observers
  (JAAVSO), 40, 430

\bibitem[{{Henry} {et~al.}(2000){Henry}, {Marcy}, {Butler}, \&
  {Vogt}}]{henry00}
{Henry}, G.~W., {Marcy}, G.~W., {Butler}, R.~P., \& {Vogt}, S.~S. 2000, \apjl,
  529, L41

\bibitem[{{H{\o}g} {et~al.}(2000){H{\o}g}, {Fabricius}, {Makarov}, {Urban},
  {Corbin}, {Wycoff}, {Bastian}, {Schwekendiek}, \& {Wicenec}}]{hog00}
{H{\o}g}, E., {et~al.} 2000, \aap, 355, L27

\bibitem[{{Howard} {et~al.}(2010){Howard}, {Johnson}, {Marcy}, {Fischer},
  {Wright}, {Bernat}, {Henry}, {Peek}, {Isaacson}, {Apps}, {Endl}, {Cochran},
  {Valenti}, {Anderson}, \& {Piskunov}}]{howard10b}
{Howard}, A.~W., {et~al.} 2010, \apj, 721, 1467

\bibitem[{{Kov{\'a}cs} {et~al.}(2005){Kov{\'a}cs}, {Bakos}, \&
  {Noyes}}]{kovacs05}
{Kov{\'a}cs}, G., {Bakos}, G., \& {Noyes}, R.~W. 2005, \mnras, 356, 557

\bibitem[{{Kov{\'a}cs} {et~al.}(2002){Kov{\'a}cs}, {Zucker}, \&
  {Mazeh}}]{kovacs02}
{Kov{\'a}cs}, G., {Zucker}, S., \& {Mazeh}, T. 2002, \aap, 391, 369

\bibitem[{{Kozai}(1962)}]{kozai62}
{Kozai}, Y. 1962, \aj, 67, 591

\bibitem[{{Lidov}(1962)}]{lidov62}
{Lidov}, M.~L. 1962, \planss, 9, 719

\bibitem[{{Martin} {et~al.}(2005){Martin}, {Fanson}, {Schiminovich},
  {Morrissey}, {Friedman}, {Barlow}, {Conrow}, {Grange}, {Jelinsky},
  {Milliard}, {Siegmund}, {Bianchi}, {Byun}, {Donas}, {Forster}, {Heckman},
  {Lee}, {Madore}, {Malina}, {Neff}, {Rich}, {Small}, {Surber}, {Szalay},
  {Welsh}, \& {Wyder}}]{martin05b}
{Martin}, D.~C., {et~al.} 2005, \apjl, 619, L1

\bibitem[{{Mason} {et~al.}(2001){Mason}, {Wycoff}, {Hartkopf}, {Douglass}, \&
  {Worley}}]{mason01}
{Mason}, B.~D., {Wycoff}, G.~L., {Hartkopf}, W.~I., {Douglass}, G.~G., \&
  {Worley}, C.~E. 2001, \aj, 122, 3466

\bibitem[{{Mayor} \& {Queloz}(1995)}]{mayor95}
{Mayor}, M., \& {Queloz}, D. 1995, \nat, 378, 355

\bibitem[{{McCullough} {et~al.}(2005){McCullough}, {Stys}, {Valenti},
  {Fleming}, {Janes}, \& {Heasley}}]{mccullough05}
{McCullough}, P.~R., {Stys}, J.~E., {Valenti}, J.~A., {Fleming}, S.~W.,
  {Janes}, K.~A., \& {Heasley}, J.~N. 2005, \pasp, 117, 783

\bibitem[{{Neveu-VanMalle} {et~al.}(2014){Neveu-VanMalle}, {Queloz},
  {Anderson}, {Charbonnel}, {Collier Cameron}, {Delrez}, {Gillon}, {Hellier},
  {Jehin}, {Lendl}, {Maxted}, {Pepe}, {Pollacco}, {S{\'e}gransan}, {Smalley},
  {Smith}, {Southworth}, {Triaud}, {Udry}, \& {West}}]{neveu14}
{Neveu-VanMalle}, M., {et~al.} 2014, \aap, 572, A49

\bibitem[{{Ochsenbein} {et~al.}(2000){Ochsenbein}, {Bauer}, \&
  {Marcout}}]{ochsenbein00}
{Ochsenbein}, F., {Bauer}, P., \& {Marcout}, J. 2000, \aaps, 143, 23

\bibitem[{{Ohta} {et~al.}(2005){Ohta}, {Taruya}, \& {Suto}}]{ohta05}
{Ohta}, Y., {Taruya}, A., \& {Suto}, Y. 2005, \apj, 622, 1118

\bibitem[{{Pejcha} {et~al.}(2013){Pejcha}, {Antognini}, {Shappee}, \&
  {Thompson}}]{pejcha13}
{Pejcha}, O., {Antognini}, J.~M., {Shappee}, B.~J., \& {Thompson}, T.~A. 2013,
  ArXiv e-prints

\bibitem[{{Pepper} {et~al.}(2007){Pepper}, {Pogge}, {DePoy}, {Marshall},
  {Stanek}, {Stutz}, {Poindexter}, {Siverd}, {O'Brien}, {Trueblood}, \&
  {Trueblood}}]{pepper07}
{Pepper}, J., {et~al.} 2007, \pasp, 119, 923

\bibitem[{{Pepper} {et~al.}(2013){Pepper}, {Siverd}, {Beatty}, {Gaudi},
  {Stassun}, {Eastman}, {Collins}, {Latham}, {Bieryla}, {Buchhave}, {Jensen},
  {Manner}, {Penev}, {Crepp}, {Cargile}, {Dhital}, {Calkins}, {Esquerdo},
  {Berlind}, {Fulton}, {Street}, {Ma}, {Ge}, {Wang}, {Mao}, {Richert}, {Gould},
  {DePoy}, {Kielkopf}, {Marshall}, {Pogge}, {Stefanik}, {Trueblood}, \&
  {Trueblood}}]{pepper13}
---. 2013, \apj, 773, 64

\bibitem[{{Perryman} {et~al.}(1997){Perryman}, {Lindegren}, {Kovalevsky},
  {Hoeg}, {Bastian}, {Bernacca}, {Cr{\'e}z{\'e}}, {Donati}, {Grenon},
  {Grewing}, {van Leeuwen}, {van der Marel}, {Mignard}, {Murray}, {Le Poole},
  {Schrijver}, {Turon}, {Arenou}, {Froeschl{\'e}}, \& {Petersen}}]{perryman97}
{Perryman}, M.~A.~C., {et~al.} 1997, \aap, 323, L49

\bibitem[{{Perryman} {et~al.}(2001){Perryman}, {de Boer}, {Gilmore}, {H{\o}g},
  {Lattanzi}, {Lindegren}, {Luri}, {Mignard}, {Pace}, \& {de
  Zeeuw}}]{perryman01}
---. 2001, \aap, 369, 339

\bibitem[{{Queloz} {et~al.}(2000){Queloz}, {Eggenberger}, {Mayor}, {Perrier},
  {Beuzit}, {Naef}, {Sivan}, \& {Udry}}]{queloz00}
{Queloz}, D., {Eggenberger}, A., {Mayor}, M., {Perrier}, C., {Beuzit}, J.~L.,
  {Naef}, D., {Sivan}, J.~P., \& {Udry}, S. 2000, \aap, 359, L13

\bibitem[{{Ricker} {et~al.}(2010){Ricker}, {Latham}, {Vanderspek}, {Ennico},
  {Bakos}, {Brown}, {Burgasser}, {Charbonneau}, {Clampin}, {Deming}, {Doty},
  {Dunham}, {Elliot}, {Holman}, {Ida}, {Jenkins}, {Jernigan}, {Kawai},
  {Laughlin}, {Lissauer}, {Martel}, {Sasselov}, {Schingler}, {Seager},
  {Torres}, {Udry}, {Villasenor}, {Winn}, \& {Worden}}]{ricker10}
{Ricker}, G.~R., {et~al.} 2010, in Bulletin of the American Astronomical
  Society, Vol.~42, American Astronomical Society Meeting Abstracts 215, 450.06

\bibitem[{{Sato} {et~al.}(2005){Sato}, {Fischer}, {Henry}, {Laughlin},
  {Butler}, {Marcy}, {Vogt}, {Bodenheimer}, {Ida}, {Toyota}, {Wolf}, {Valenti},
  {Boyd}, {Johnson}, {Wright}, {Ammons}, {Robinson}, {Strader}, {McCarthy},
  {Tah}, \& {Minniti}}]{sato05}
{Sato}, B., {et~al.} 2005, \apj, 633, 465

\bibitem[{{Schlegel} {et~al.}(1998){Schlegel}, {Finkbeiner}, \&
  {Davis}}]{schlegel98}
{Schlegel}, D.~J., {Finkbeiner}, D.~P., \& {Davis}, M. 1998, \apj, 500, 525

\bibitem[{{Schneider} {et~al.}(2011){Schneider}, {Dedieu}, {Le Sidaner},
  {Savalle}, \& {Zolotukhin}}]{schneider11}
{Schneider}, J., {Dedieu}, C., {Le Sidaner}, P., {Savalle}, R., \&
  {Zolotukhin}, I. 2011, \aap, 532, A79

\bibitem[{{Seager} \& {Mall{\'e}n-Ornelas}(2003)}]{seager03}
{Seager}, S., \& {Mall{\'e}n-Ornelas}, G. 2003, \apj, 585, 1038

\bibitem[{{Siess} {et~al.}(2000){Siess}, {Dufour}, \& {Forestini}}]{siess00}
{Siess}, L., {Dufour}, E., \& {Forestini}, M. 2000, \aap, 358, 593

\bibitem[{{Siverd} {et~al.}(2012){Siverd}, {Beatty}, {Pepper}, {Eastman},
  {Collins}, {Bieryla}, {Latham}, {Buchhave}, {Jensen}, {Crepp}, {Street},
  {Stassun}, {Gaudi}, {Berlind}, {Calkins}, {DePoy}, {Esquerdo}, {Fulton},
  {Furesz}, {Geary}, {Gould}, {Hebb}, {Kielkopf}, {Marshall}, {Pogge},
  {Stanek}, {Stefanik}, {Szentgyorgyi}, {Trueblood}, {Trueblood}, {Stutz}, \&
  {van Saders}}]{siverd12}
{Siverd}, R.~J., {et~al.} 2012, ArXiv e-prints

\bibitem[{{Skrutskie} {et~al.}(2006){Skrutskie}, {Cutri}, {Stiening},
  {Weinberg}, {Schneider}, {Carpenter}, {Beichman}, {Capps}, {Chester},
  {Elias}, {Huchra}, {Liebert}, {Lonsdale}, {Monet}, {Price}, {Seitzer},
  {Jarrett}, {Kirkpatrick}, {Gizis}, {Howard}, {Evans}, {Fowler}, {Fullmer},
  {Hurt}, {Light}, {Kopan}, {Marsh}, {McCallon}, {Tam}, {Van Dyk}, \&
  {Wheelock}}]{skrutskie06}
{Skrutskie}, M.~F., {et~al.} 2006, \aj, 131, 1163

\bibitem[{{Smalley} {et~al.}(2012){Smalley}, {Anderson}, {Collier-Cameron},
  {Doyle}, {Fumel}, {Gillon}, {Hellier}, {Jehin}, {Lendl}, {Maxted}, {Pepe},
  {Pollacco}, {Queloz}, {S{\'e}gransan}, {Smith}, {Southworth}, {Triaud},
  {Udry}, \& {West}}]{smalley12}
{Smalley}, B., {et~al.} 2012, \aap, 547, A61

\bibitem[{{Spiegel} \& {Madhusudhan}(2012)}]{spiegel12}
{Spiegel}, D.~S., \& {Madhusudhan}, N. 2012, \apj, 756, 132

\bibitem[{{Steffen} {et~al.}(2012){Steffen}, {Ford}, {Rowe}, {Fabrycky},
  {Holman}, {Welsh}, {Batalha}, {Borucki}, {Bryson}, {Caldwell}, {Ciardi},
  {Jenkins}, {Kjeldsen}, {Koch}, {Pr{\v s}a}, {Sanderfer}, {Seader}, \&
  {Twicken}}]{steffen12}
{Steffen}, J.~H., {et~al.} 2012, \apj, 756, 186

\bibitem[{{Torres} {et~al.}(2010){Torres}, {Andersen}, \&
  {Gim{\'e}nez}}]{torres10}
{Torres}, G., {Andersen}, J., \& {Gim{\'e}nez}, A. 2010, \aapr, 18, 67

\bibitem[{{Torres} {et~al.}(2012){Torres}, {Fischer}, {Sozzetti}, {Buchhave},
  {Winn}, {Holman}, \& {Carter}}]{torres12}
{Torres}, G., {Fischer}, D.~A., {Sozzetti}, A., {Buchhave}, L.~A., {Winn},
  J.~N., {Holman}, M.~J., \& {Carter}, J.~A. 2012, \apj, 757, 161

\bibitem[{{Triaud} {et~al.}(2010){Triaud}, {Collier Cameron}, {Queloz},
  {Anderson}, {Gillon}, {Hebb}, {Hellier}, {Loeillet}, {Maxted}, {Mayor},
  {Pepe}, {Pollacco}, {S{\'e}gransan}, {Smalley}, {Udry}, {West}, \&
  {Wheatley}}]{triaud10}
{Triaud}, A.~H.~M.~J., {et~al.} 2010, \aap, 524, A25

\bibitem[{{van Leeuwen}(2007)}]{vanleeuwen07}
{van Leeuwen}, F. 2007, \aap, 474, 653

\bibitem[{{Vidal-Madjar} {et~al.}(2003){Vidal-Madjar}, {Lecavelier des Etangs},
  {D{\'e}sert}, {Ballester}, {Ferlet}, {H{\'e}brard}, \& {Mayor}}]{vidal03}
{Vidal-Madjar}, A., {Lecavelier des Etangs}, A., {D{\'e}sert}, J., {Ballester},
  G.~E., {Ferlet}, R., {H{\'e}brard}, G., \& {Mayor}, M. 2003, \nat, 422, 143

\bibitem[{{Vogt} {et~al.}(1994){Vogt}, {Allen}, {Bigelow}, {Bresee}, {Brown},
  {Cantrall}, {Conrad}, {Couture}, {Delaney}, {Epps}, {Hilyard}, {Hilyard},
  {Horn}, {Jern}, {Kanto}, {Keane}, {Kibrick}, {Lewis}, {Osborne},
  {Pardeilhan}, {Pfister}, {Ricketts}, {Robinson}, {Stover}, {Tucker}, {Ward},
  \& {Wei}}]{vogt94}
{Vogt}, S.~S., {et~al.} 1994, in , 362

\bibitem[{{Wang} {et~al.}(2012){Wang}, {Wright}, {Cochran}, {Kane}, {Henry},
  {Payne}, {Endl}, {MacQueen}, {Valenti}, {Antoci}, {Dragomir}, {Matthews},
  {Howard}, {Marcy}, {Isaacson}, {Ford}, {Mahadevan}, \& {von Braun}}]{wang12}
{Wang}, Sharon, X., {et~al.} 2012, \apj, 761, 46

\bibitem[{{Winn}(2010)}]{winn10a}
{Winn}, J.~N. 2010, {Exoplanet Transits and Occultations}, ed. {Seager, S.},
  55--77

\bibitem[{{Winn} {et~al.}(2010){Winn}, {Fabrycky}, {Albrecht}, \&
  {Johnson}}]{winn10c}
{Winn}, J.~N., {Fabrycky}, D., {Albrecht}, S., \& {Johnson}, J.~A. 2010, \apjl,
  718, L145

\bibitem[{{Winn} {et~al.}(2005){Winn}, {Noyes}, {Holman}, {Charbonneau},
  {Ohta}, {Taruya}, {Suto}, {Narita}, {Turner}, {Johnson}, {Marcy}, {Butler},
  \& {Vogt}}]{winn05}
{Winn}, J.~N., {et~al.} 2005, \apj, 631, 1215

\bibitem[{{Wright} {et~al.}(2010){Wright}, {Eisenhardt}, {Mainzer}, {Ressler},
  {Cutri}, {Jarrett}, {Kirkpatrick}, {Padgett}, {McMillan}, {Skrutskie},
  {Stanford}, {Cohen}, {Walker}, {Mather}, {Leisawitz}, {Gautier}, {McLean},
  {Benford}, {Lonsdale}, {Blain}, {Mendez}, {Irace}, {Duval}, {Liu}, {Royer},
  {Heinrichsen}, {Howard}, {Shannon}, {Kendall}, {Walsh}, {Larsen}, {Cardon},
  {Schick}, {Schwalm}, {Abid}, {Fabinsky}, {Naes}, \& {Tsai}}]{wright10}
{Wright}, E.~L., {et~al.} 2010, \aj, 140, 1868

\bibitem[{{Wright} {et~al.}(2011){Wright}, {Fakhouri}, {Marcy}, {Han}, {Feng},
  {Johnson}, {Howard}, {Fischer}, {Valenti}, {Anderson}, \&
  {Piskunov}}]{wright11}
{Wright}, J.~T., {et~al.} 2011, \pasp, 123, 412

\bibitem[{{Wu} \& {Murray}(2003)}]{wu03}
{Wu}, Y., \& {Murray}, N. 2003, \apj, 589, 605

\bibitem[{{Yelda} {et~al.}(2010){Yelda}, {Lu}, {Ghez}, {Clarkson}, {Anderson},
  {Do}, \& {Matthews}}]{yelda10}
{Yelda}, S., {Lu}, J.~R., {Ghez}, A.~M., {Clarkson}, W., {Anderson}, J., {Do},
  T., \& {Matthews}, K. 2010, \apj, 725, 331

\bibitem[{{Yi} {et~al.}(2001){Yi}, {Demarque}, {Kim}, {Lee}, {Ree}, {Lejeune},
  \& {Barnes}}]{yi01}
{Yi}, S., {Demarque}, P., {Kim}, Y.-C., {Lee}, Y.-W., {Ree}, C.~H., {Lejeune},
  T., \& {Barnes}, S. 2001, \apjs, 136, 417

\end{thebibliography}

\end{document}